# *NFTracer*: Tracing NFT Impact Dynamics in Transaction-flow Substitutive Systems with Visual Analytics

Yifan Cao , Qing Shi , Lue Shen , Kani Chen , Yang Wang , Wei Zeng  *Member, IEEE*, Huamin Qu  *Member, IEEE*

**Abstract**—Impact dynamics are crucial for estimating the growth patterns of NFT projects by tracking the diffusion and decay of their relative appeal among stakeholders. Machine learning methods for impact dynamics analysis are incomprehensible and rigid in terms of their interpretability and transparency, whilst stakeholders require interactive tools for informed decision-making. Nevertheless, developing such a tool is challenging due to the substantial, heterogeneous NFT transaction data and the requirements for flexible, customized interactions. To this end, we integrate intuitive visualizations to unveil the impact dynamics of NFT projects. We first conduct a formative study and summarize analysis criteria, including substitution mechanisms, impact attributes, and design requirements from stakeholders. Next, we propose the *Minimal Substitution Model* to simulate substitutive systems of NFT projects that can be feasibly represented as node-link graphs. Particularly, we utilize attribute-aware techniques to embed the project status and stakeholder behaviors in the layout design. Accordingly, we develop a multi-view visual analytics system, namely *NFTracer*, allowing interactive analysis of impact dynamics in NFT transactions. We demonstrate the informativeness, effectiveness, and usability of *NFTracer* by performing two case studies with domain experts and one user study with stakeholders. The studies suggest that NFT projects featuring a higher degree of similarity are more likely to substitute each other. The impact of NFT projects within substitutive systems is contingent upon the degree of stakeholders' influx and projects' freshness.

✦

## 1 INTRODUCTION

INTEREST in non-fungible tokens (NFTs) has surged globally in recent years. In brief, an NFT is a virtual token that exists on a blockchain and can be paired with diverse digital or physical assets to verify assets' *ownership* and *authenticity* [1], [2], [3]. For instance, NFTs could be applied to the art industry to cryptographically certify the *provenance* of artworks by storing the metadata concerning collection and transaction activities on the blockchain [4], [5]. With such technical and economic merits, tons of enthusiasts have flooded NFT marketplaces as different stakeholders [6], [7]. These stakeholders can be broadly categorized into four main groups: creators, collectors, moderators, and developers [8]. Among these groups, collectors and investors, especially those with mathematics, finance, and business backgrounds, are the most active and numerous within the marketplace [9]. They often engage in a high volume of transactions. These stakeholders believe NFTs have been gradually developed into an alternative investment, specif-

• *Yifan Cao, Lue Shen, Kani Chen, Yang Wang, and Huamin Qu are with the Hong Kong University of Science and Technology. E-mails: {ycaoaw, lucas.shen}@connect.ust.hk, {makchen, yangwang}@ust.hk, and huamin@cse.ust.hk.*
• *Qing Shi was with the Hong Kong University of Science and Technology (Guangzhou). E-mails: brantqshi@hkust-gz.edu.cn.*
• *Wei Zeng is with the Hong Kong University of Science and Technology (Guangzhou) and the Hong Kong University of Science and Technology. Email: weizeng@hkust-gz.edu.cn.*
• *Wei Zeng is the corresponding author.*

ically in decentralized finance (DeFi) and the creative industry [10], [11]. Based on a rough estimate conducted in 2022, over 265,000 active wallets are trading NFTs on the Ethereum blockchain alone [12].

Nevertheless, the lack of assessment metrics for NFTs has long been a pain point that jeopardizes the development of NFT marketplaces [8]. Take NFT projects, i.e., collections containing a series of NFT collectibles, as an example. One of the most common investment risks is the sudden collapse of NFT projects because they cannot keep up with rapid iteration and fall out of market competition. Stakeholders who invest in such NFT projects may face substantial losses due to misjudging a project's lifespan. Moreover, stakeholders suffer from other risks, such as the irrational volatility of NFT prices, phishing scams [13], hype [14], wash trading [4], and "rug pull" [15], that could abruptly devalue and interrupt the impact of NFT projects. Detecting the **impact dynamics** is crucial for stakeholders to estimate the growth pattern and anticipate the relative appeal of NFT projects. More broadly, impact dynamics represent the diffusion and decay process of an object's appeal to its target users, such as commercial products, fashion trends, and techniques [16]. Although traditional machine learning models can predict the impact dynamics of NFT projects, they lack interpretation and cannot accommodate stakeholders with diverse requirements. Thereby, a comprehensive analytical tool for detecting the impact dynamics of NFT projects is needed. Such a tool would be particularly beneficial for stakeholders with STEM (Science, Technology, Engineering, and Mathe-matics) backgrounds skilled in using digital dashboards to aid their decision-making processes.

To bridge this research gap, we integrate visualization techniques to facilitate the impact dynamics analysis of NFT transactions. Specifically, we propose the *Minimal Substitution (MS) Model* to simulate and exhibit **substitutive systems** that illustrate how stakeholders migrate among different projects and thus reflect their relative appeal [17], [18]. However, we encounter three challenges during implementation:

**First, identifying substitution mechanisms and impact attributes is non-trivial.** Deriving the mechanisms that govern the substitutive systems out of multiple alternatives is challenging. Even with identified mechanisms, refining the underlying impact attributes from substantial and heterogeneous transaction data, such as historical market performance, social media activities, and transaction behaviors, requires a detailed investigation [8], [19].

**Second, visualizing substitutive systems and impact dynamics for stakeholders from diverse backgrounds is demanding.** It requires providing an intuitive and well-coordinated design for multivariate and voluminous NFT transaction data, necessitating sophisticated data transformation, effective database management, and a comprehensive and multi-dimensional approach.

**Third, fulfilling the personalized analysis requirements of diversified stakeholders presents a challenge.** This complexity arises because NFT projects have varying launch and extinction dates, which different stakeholders may find relevant. Additionally, stakeholders with diverse market roles often prefer different impact attributes [6], [7]. For instance, while moderators may focus more on social media factors, investors typically prioritize floor prices and sales volumes.

For the first challenge, we conduct a formative study with domain experts and stakeholders to identify analysis criteria. Accordingly, we propose the *MS Model* to simulate the substitutive systems in NFT transactions. For the second challenge, we leverage a visual analytics (VA) approach to present the substitutive systems of NFT transactions for analyzing NFT impact dynamics. We metaphorically showcase the substitutive systems via an augmented node-link graph with multi-attribute-aware techniques [20], [21] to embed the project status and stakeholder behaviors in the layout design. We apply K-means and the Gaussian mixture model to cluster similar NFT projects and improve the efficiency of users' workflows. To tackle the third challenge, we develop *NFTracer*, an interactive, well-coordinated VA system to provide hierarchical explorations primarily for stakeholders with STEM backgrounds. In addition, *NFTracer* also aspires to inspire stakeholders from humanities or social science backgrounds to understand and analyze NFT marketplaces from a systematic perspective.

The main contributions of this paper are as follows:

- Proposing the *Minimal Substitution (MS) Model* to simulate substitutive systems in NFT transactions with underlying substitution mechanisms and impact attributes characterized.
- Developing *NFTracer*, which, to our knowledge, is the first interactive VA system for different stakeholders to explore the substitutive systems and impact dynamics of NFT projects.

TABLE 1: The key concepts for detecting substitutive systems in NFT transactions for impact dynamics analysis.

| Concepts | Explanation |
|---|---|
| NFT Stakeholders | Participants playing different roles in the NFT marketplaces, e.g., investors, buyers, sellers, and holders. |
| NFT Projects | A collection of NFT assets containing a limited number of similarly styled NFTs with minor modifications in design. |
| Substitution Flows | Flows consisting of migrated stakeholders among NFT projects with direction and intensity encoded. |
| Substitutive Systems | A node-link graph containing NFT projects connected by substitution flows, displaying users' inflow and outflow tendency. |
| Substitution Mechanisms | Generic ingredients governing the probability of substitution, i.e., substitutive rate between NFT projects in our case. |
| Impact Attributes | Factors influencing the mechanisms governing substitutive systems. |
| Impact Dynamics | The evolution of the global appeal of individual NFT projects to stakeholders in substitutive systems. |

- Evaluating *NFTracer* through two case studies with domain experts to derive in-depth insights and conducting a user study with 13 stakeholders to demonstrate the effectiveness of our system.

## 2 BACKGROUND AND RELATED WORK

This section introduces the key concepts for detecting NFT substitutive systems (Tab. 1), followed by related work categorized as *impact dynamics in substitutive systems*, *NFT assessment analysis*, and *network analysis of NFT communities*.

### 2.1 Impact Dynamics in Substitutive Systems

**Impact** is a "*brand-level*" [22], [23] parameter that evaluates the comparative attractiveness and visibility of analogous products [16], [24], services [23], [25], information [26], [27], knowledge [28], [29], opinions [30], [31], and technologies [32] within their respective social systems. As such, the **impact dynamic** reflects the progression of impact over time, encapsulating the competition in adoption and loyalty towards a commodity, opinion, or technology among a target population in substitutive systems [16], [17], [33]. Specifically, impact dynamics metrics are obtained from studies focused on modeling **innovation diffusion** [34] of new-product growth in competitive market environments [35]. Among the foremost and crucial models in this field is the Bass Diffusion Model [36]. This model delineates the S-shaped pattern of new-product growth and presents three key parameters—internal influences, external influences, and market potential [34], [37]. Based on the Bass model's assumptions, subsequent researchers have refined and broadened its parameters by exploring case studies in various industries and domains [26], [29], [31].

Particularly, accelerated product substitution has piqued scholars' interest in the study of **customer attrition** [16], [28], e.g., brand-shift, in competitive markets. Customer attrition happens in areas with a finite customer base or sluggish growth, such as the contemporary NFT ecosystem [38]. This often results in competition for finite customers. Consequently, numerous new products (e.g., emerging NFT projects) struggle to surmount the Market Chasm [39] and fulfill their anticipated lifecycle, potentially giving rise to a vicious cycle of market development and squandering of wealth and resources [32], [40]. Hence, stakeholders need to assess impact dynamics in rapidly changing substitutive systems like the NFT ecosystem to forecast a product's longevity and market competitiveness [14], [17].

TABLE 2: Definitions for mechanisms, impact attributes, and related data to construct NFT substitutive systems.

| Mechanisms | Definition | Corresponding Impact Attributes | Related Data |
|---|---|---|---|
| M1: Recency | The anticipated longevity of NFT projects or, simply put, how new NFT projects are. | Time stamps; Social media popularity. | Social media data. |
| M2: Preferential Attachment | The extent to which a certain NFT project is more likely to attract new collectors than less popular counterparts. | Transaction behaviors of stakeholders, indicated by the number of sellers, buyers, and holders. | Transaction data; Project character data. |
| M3: Propensity | The extent to which a certain NFT project is more likely to substitute for some NFT projects than others. | Inherent characteristics of NFT projects, e.g., floor price, number of whales, sales volume, etc. | Project character data; Social media data. |

## 2.2 NFT Assessment Analysis

Non-fungible tokens (NFTs) are digital certificates derived from blockchain technology for which smart contracts generate and verify automatically [41]. Characterized as "*unique, indivisible, irreplaceable, and verifiable*", NFTs provide new authentication methods and enable personal ownership of scarce digital properties, e.g., digital art, collectibles, and virtual land in the Metaverse [42]. In the fourth quarter of 2022, NFT trading platforms achieved a market capitalization of $22.56 billion within just two years, accounting for 18% of the entire crypto industry [43]. Recently, a commercial report published by Statista predicts that the number of stakeholders participating in NFT marketplaces is expected to reach approximately 19.31 million by 2027 [44]. Currently, NFTs have become the alternative investment and have hastened various marketplaces [4], [10], e.g., OpenSea [45], Rarible [8], and SuperRare [19].

However, the lack of assessment metrics plagues stakeholders and hinders the development of NFT marketplaces [8]. Since 2021, *NFT assessment analysis* has accumulated numerous insights into the value indicators of NFTs. Such indicators contain two dimensions: 1) *intrinsic influential factors*, e.g., historical transaction records and visual components [19], [46], [47], [48], and 2) *extrinsic relevant factors*, e.g., bitcoin volatility and media exposure [49], [50], [51], [52]. For instance, Nadini *et al.* [19] distilled the influential statistical properties of NFT sales and trained a machine learning model to predict secondary sales volume. Yousaf *et al.* [46] further inspected the interrelationship among intrinsic influential factors, i.e., trading volume, volatility, and returns, of NFTs to facilitate investment decision-making. In addition, extant literature also estimates the efficacy of extrinsic factors through correlation analysis and deep learning models. Respectively, Kapoor *et al.* [51] and Wilkoff *et al.* [52] confirmed that social media popularity and media reports could impact the liquidity and price of NFTs.

Nevertheless, the above insights are mainly drawn from intricate machine learning models with fixed variables, which are too incomprehensible and rigid for general users to understand. Hence, we conducted a mixed-method formative study with domain experts and stakeholders to summarize their design requirements to ensure interpretability and transparency. Accordingly, we distill analysis criteria and present the *MS Model* for simulating substitutive systems of NFT transactions. Then, we employ the VA system, *NFTracer*, as one of the first tools for stakeholders to explore the impact dynamics of NFT projects.

## 2.3 Network Analysis of NFT Communities

Another major research focus involves analyzing stakeholder behaviors within NFT communities by visualizing their dynamic social networks [53], [54].

Previous investigations can be classified into three progressive dimensions: 1) *identifying stakeholders* [6], [7], 2) *understanding stakeholders' interactions* [8], and 3) *constructing graph model for stakeholders' activities* [53], [55]. To start with, Baytaş *et al.* [7] proposed a model mapping six salient stakeholders and their relations in NFT marketplaces through content analysis on a social media website. Then, the interactions between NFT stakeholders have drawn attention and inspired fine-grained network analysis on an address basis for identifying stakeholders' transaction patterns. For example, Vasan *et al.* [54] characterized the homophily pattern of NFT communities, composed mainly of artists and their collectors, via network analysis. Brunet *et al.* [55] further uncovered the guiding roles played by successful investors and big accounts by extracting the topological structure of NFT transaction networks. Besides, Wen *et al.* [56] applied a visual analytics approach to expose the wash trading networks of NFT flows in collectors' communities.

Although the aforementioned network visualizations contribute to uncovering transaction patterns and abnormal behaviors, visual clutter is almost inevitable when including enormous addresses in analysis. Moreover, their layout designs are limited in visually presenting fine-grained transaction behaviors of stakeholders or allowing users to directly connect transaction patterns with influential factors. Such scalability issues were also common in network analysis concerning transactions in the broader blockchain ecosystem, e.g., Bitcoin exchanges and mining pools [57], [58], [59]. To address this research gap, we use two sparse configuration techniques to improve the readability of node-link graphs. In addition, we apply multi-attribute-aware techniques [20], [21] to embed the project status and stakeholder behaviors in the layout for straightforward comparison.

## 3 USER-CENTERED DESIGN

We conducted a mixed-method formative study to identify substitution mechanisms, impact attributes, and design requirements from domain experts and stakeholders (Fig. 1).

### 3.1 Focus Group and Mechanisms

We involved four domain experts for over six months ($E_{A-D}$) to collaboratively summarize the analysis criteria of substitutive systems in NFT transactions. $E_A$ and $E_B$ are scholars investigating stakeholders' behaviors in NFT marketplaces. $E_C$ is a senior investor focusing on NFT projects. $E_D$ is a professional researcher dedicated to quantifying and optimizing regulation strategies in NFT communities. We organized a two-hour focus group with three domain experts to define the research scope (Fig. 1A). Prior to the discussion, we prepared seven potential substitution models (please refer to Tab. A1 in Appendix A.1 for details), classified based on their primary application contexts and provided corresponding substitution mechanisms to aid the experts' discussion [24], [33], [60], [61], [62], [63], [64]. To foster an efficient discussion among domain experts, we disseminated the list of candidate models to them one week before the group discussion, allowing the experts to familiarize themselves with the material.

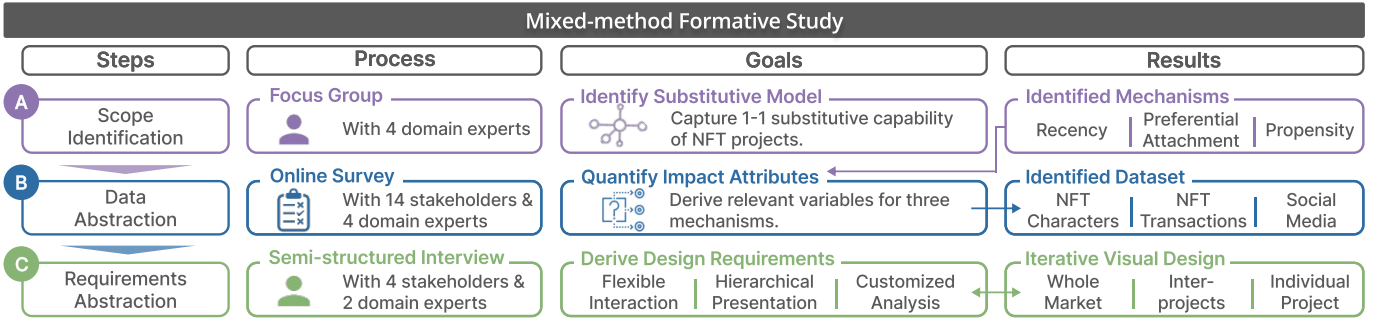

Fig. 1: The pipeline of conducting the formative study with domain experts and NFT stakeholders to identify the research scope, analysis criteria (i.e., mechanisms and impact attributes), related data, and design requirements.

In the initial 50-minute session, we presented the foundational assumptions, relevant use cases, and key mechanisms of candidate models to domain experts. We discovered that $E_A$ and $E_B$, proficient in qualitative research, encountered difficulties directly evaluating and selecting candidate models. $E_A$ remarked, "*These candidates indicate that similarities influence the impact dynamics of NFT projects, reflecting the **propensity** or **fitness** mechanism. However, I need more information to understand and compare other mechanisms.*"

Thus, we took a 30-minute break and then added extant literature focusing on quantifying the market performance of NFT projects to support their discussion and decision-making in the second session [7], [19], [41], [52], [65]. For the next 40 minutes, domain experts compared and analyzed key mechanisms, ultimately excluding four candidate models. These models were either overly complex (e.g., *Lotka-Volterra Competition Model*), focused too heavily on graph data structures without addressing any domain knowledge (e.g., the *Evolving Network Model*), or incompatible with the characteristics of NFT trading (e.g., the *Susceptible-Infected-Recovered Model*). Hence, domain experts narrowed down candidate models to the *Bass Diffusion Model* (C1), *Minimal Substitution (MS) Model* (C5), and *Gompertz Model* (C7) through voting. Notably, $E_C$ stated, "*Unlike the resolution of illnesses, users in the NFT market seldom exit completely. Instead, they migrate among similar NFT projects. This phenomenon aligns with the mechanism of **preferential attachment**.*"

After another break, we presented the selected three candidate models with demo codes (see Appendix A.2 for model fitting results) and then invited domain experts to rank their applicability. Specifically, $E_A$ and $E_D$ emphasized the significant impact of release timing on the appeal of NFT projects. $E_D$ explained, "*In the NFT ecosystem, stakeholders find it difficult to develop loyalty to projects, making the initial launch period the most appealing. Although **recency** is a straight-*

*forward mechanism, it is essential for measuring the allure of fast-paced products like NFTs.*"

Finally, the focus group reached a consensus on the *Minimal Substitution (MS) Model* and its three major mechanisms: 1) *recency*, 2) *preferential attachment*, and 3) *propensity* (Tab. 2) to simulate NFT substitutive systems [16]. Domain experts discussed impact attributes for quantifying each attribute's impact degree on mechanisms but could not agree on each attribute's impact degree on mechanisms. In this light, a subsequent online survey was conducted with NFT stakeholders.

### 3.2 Online Survey and Impact Attributes

The online survey contains three subsections: 1) briefing to participants about the research topic and three mechanisms; 2) collecting consent forms and demographic information; and 3) asking stakeholders to evaluate 15 potential attributes with a five-point Likert scale (Fig. 1B). Besides the domain experts, we recruited another 14 stakeholders from NFT communities. Ultimately, we received 18 responses (P1-18: M=11, F=7), which included five investors, three researchers, three collectors, and seven respondents with multiple identities (Please refer to Tab. A3 in Appendix A.3). Their ages ranged from 25 to 36 (Avg = 27.39, SD = 3.05), and their engagement in NFT marketplaces extended from six months to five years (Avg = 2.28, SD = 1.44). We used parametric methods [66] to assess how each potential attribute affected the individual mechanisms. According to the results (please refer to Fig. A2 in Appendix A.4), we summarized 12 fundamental impact attributes and constructed our database (Fig. 2A1-2). We found that domain experts and stakeholders generally agreed on impact attributes for quantifying *preferential attachment* and *recency*. However, multiple impact attributes were assessed as relevant for *propensity*, which led us to conduct semi-structured interviews to better understand users' requirements.

### 3.3 Interviews and Design Requirements

We conducted semi-structured interviews with four stakeholders and two domain experts (I1-6) to extract their design requirements for NFT impact dynamics analysis (Fig. 1C). Two authors conducted a thematic analysis [67] iteratively based on responses until the inter-rater reliability reached 0.81, indicating that the two coders reached a satisfactory agreement. Accordingly, we distilled six requirements. First, I1 and I3 suggested leveraging *flexible interactions* in the VA system, as they prefer to choose relevant data indicators and make judgments based on their past trading experiences. For instance, they opposed pre-slicing time windows for target users, which may unreasonably interrupt some projects' life cycles and create exploration barriers (R1). Meanwhile, they recommended visually presenting the quantified mechanisms to better interpret the *MS Model* (R2). In addition, I4 and I5 highlighted *customized analysis*, particularly when quantifying the *propensity* for different stakeholders (R3). They also propounded filtering and comparing specific NFT projects based on their mechanism distribution (R4). Lastly, I2 and I6 appreciated *hierarchical presentation*. More specifically, they would like to observe the whole substitutive systems, how similar NFT projects substitute each other, and finally, the impact dynamics of individual NFT projects (R5). Furthermore, they proposed that comparisons of impact dynamics and growth patterns between different NFT projects would be beneficial for their analysis (R6).

TABLE 3: Bridging categorized visualization tasks with corresponding design requirements of domain experts.

| Category | | Visualization Tasks | Design Requirements | System View |
|---|---|---|---|---|
| Overview | T1 | To visualize the overall substitutive networks in NFT transactions in selected duration. | R1 | Substitution View |
| | T2 | To present the distribution of three mechanisms and impact dynamics on a group basis. | R2, R4 | Mechanisms Analysis View |
| Comparison | T3 | To support inter-cluster comparisons of shared characteristics among grouped NFT projects. | R2, R3, R4 | Mechanisms Analysis View |
| | T4 | To support intra-cluster comparisons of stakeholders' interactions among individual NFT projects. | R5, R6 | Preferential Attachment View |
| Clustering | T5 | To support cluster analysis based on personalized attribute selection. | R1, R3 | Propensity Analysis View |
| Temporal Trend Analysis | T6 | To display the temporal evolution of substitutive systems in NFT transactions. | R1 | Substitution View |
| | T7 | To present the temporal impact dynamics of individual NFT projects. | R5, R6 | Impact Dynamic View |
| Feature Extraction | T8 | To distill and quantify three mechanisms from heterogeneous transaction data. | R2, R4 | Mechanisms Analysis View |
| | T9 | To simulate the node-link graph for substitutive networks based on extracted mechanisms. | R1 | Substitution View |

### 3.4 Requirement Analysis

We summarized and categorized the six design requirements into three aspects to establish hierarchical and well-coordinated views for users to explore the substitutive systems in NFT transactions.

**Market-level Interpretation** enables flexible analysis of the overall substitutive systems of NFT transactions.

**R1. Provide personalized demonstration of the overall substitutive systems in NFT transactions.** NFT projects vary in launch date and duration. An overview of substitutive systems with flexible time-slicing options is thus necessary for stakeholders' analysis.

**R2. Quantify mechanisms governing substitutive systems in NFT transactions.** Stakeholders, especially those unfamiliar with data analysis or substitution models, require an intuitive mapping of quantified mechanisms and impact attributes. Hence, a broader range of users can fully understand the substitution of NFT projects.

**Inter-project level Investigation** provides tailored analysis to discern similar NFT projects in substitutive systems.

**R3. Recommend similar NFT projects with higher substitutive probability.** Detecting similarities between NFT projects can assist stakeholders in filtering those with a higher probability of substitution. The similarity levels are mainly measured by the *propensity* mechanism that considers multiple attributes, which are characterized according to users' requirements.

**R4. Compare patterns of substitution mechanisms and impact dynamics of grouped NFT projects.** Stakeholders require identifying patterns of mechanisms to summarize common characteristics of NFT projects that fall into the same cluster and compare NFT projects belonging to different clusters within a selected duration.

**Project-level Exploration** facilitates hierarchical and fine-grained analysis of growth patterns and impact dynamics within individual NFT projects in substitutive systems.

**R5. Capture impact dynamics of individual NFT projects over time.** Impact dynamics in substitutive systems reflect the appeal and vitality of individual NFT projects, which helps stakeholders to anticipate their longevity. Thus, this information is helpful for stakeholders when making decisions about whether to hold or abandon NFT projects.

**R6. Compare impact dynamics and growth patterns between NFT projects.** The market performance of NFT projects fluctuates dramatically. Hence, enabling users to compare the impact dynamics and growth patterns of NFT projects to react quickly is crucial.

## 4 VISUALIZATION TASKS AND SYSTEM OVERVIEW

This section introduces the visualization tasks based on abstracted requirements introduced in Sec. 3.4 and provides a system overview to illustrate the whole framework.

### 4.1 Visualization Tasks

We identify nine visualization tasks (**T1-9**) based on the design requirements concluded from the formative study (see Tab. 3). These tasks are structured following the classification of visual analytic tasks for ensemble data proposed by Wang *et al.* [68]. **T1** and **T2** aim to provide a visual overview of the three mechanisms and the overall topology of substitutive systems in NFT transactions. **T3** and **T4** discern the distinctions between clustered and individual NFT projects through juxtaposition and superimposition layouts. **T5** categorizes NFT projects based on their resemblances. **T6** and **T7** uncover the temporal development of individual NFT projects, while **T8** and **T9** extract the underlying mechanisms from latent impact attributes that influence substitutive systems in NFT transactions. In the following sections, we primarily use the six "design requirements" to map visual designs for brevity.

### 4.2 System Overview

*NFTracer* is a web-based VA system consisting of three modules: *Data Storage*, *Data Analysis*, and *Data Visualization* (Fig. 2). The *Data Storage* module is a database storing collected and transformed numerical tabular data of NFT transaction records, project characters, and social media content. We reorganized these data into indexed impact attributes and created attribute-related relationship tables to provide flexible and efficient interactions. The *Data Analysis* module supports a personalized selection of time windows and impact attributes, returning corresponding quantified mechanisms, substitution flows, and impact dynamics to users on demand. These two modules are implemented in Python and Flask and form the backend of *NFTracer*. The *Data Visualization* module utilized Vue.js and D3.js [69] to build a frontend application with four well-coordinated views that support interactive explorations.

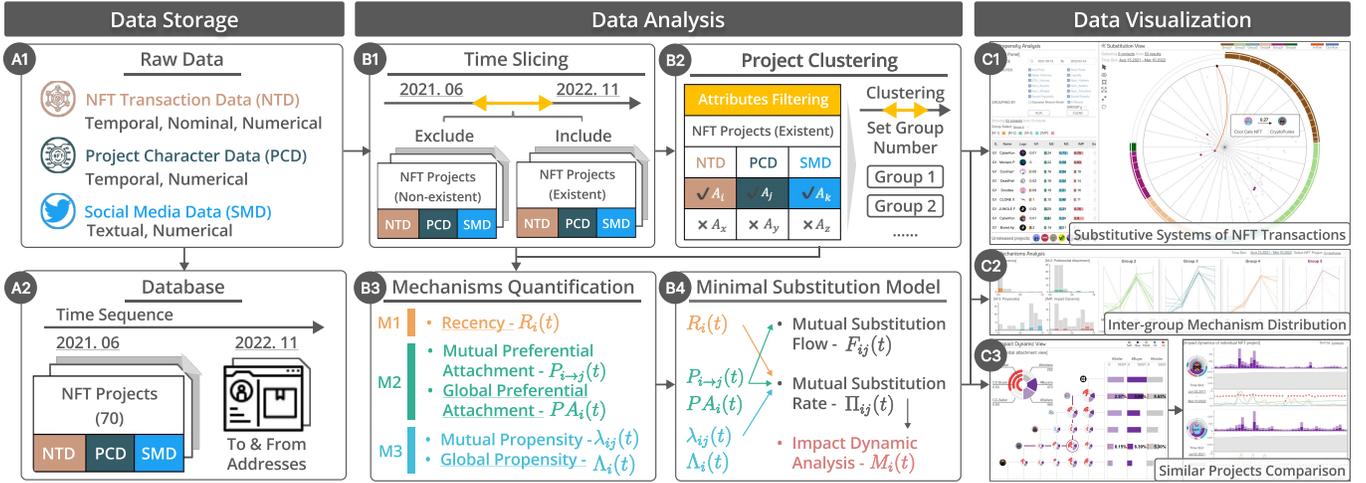

Fig. 2: System Overview. *NFTracer* comprises three modules, which are *Data Storage* (A1-2), *Data Analysis* (B1-4) for clustering NFT projects and quantifying mechanisms with personalized time slicing, and *Data Visualization* (C1-3).

## 5 DATA PROCESSING

This section describes the input data, outlines the data pre-processing pipeline, and explains the measurement of three substitution mechanisms and other parameters defining NFT substitutive systems. To facilitate comparison, we apply the Min-Max normalization technique to ensure the results of each step of data processing fall within [0,1].

### 5.1 Data Description

The database for exploring substitutive systems in NFT transactions is heterogeneous, containing data manually collected from three sources, i.e., Etherscan [70], NFTGO [9], and Twitter. We discussed with four domain experts to determine sampling methods for NFT projects. They noted that many NFT projects have a short lifespan, minimal trading volume, and generate low economic value, which makes them unqualified for impact analysis. Accordingly, we took market performance (i.e., market capitalization and sales volumes) and length of existence as our selection criteria. Based on this, we sampled 70 NFT projects that have been ranked as top-tier projects on OpenSea from June 2021 to November 2022 (see Tab. B4 in Appendix B). Specifically, the combined market cap of the 70 sampled NFT projects represents 50.81% of the total market cap of all NFT projects (n = 23,329). Additionally, the combined sales volumes of these sampled projects account for 51.81% of the total volume. These figures indicate that our sampling is representative and the trading and migration activities within these NFT projects can provide insights into the overall situation of NFT marketplaces. Then, we applied a web crawler and API acquisition method for data collection by searching the names, smart contract addresses, or Twitter hashtags of the selected NFT projects. As such, we gathered *NFT transaction data*, *project character data*, and *social media data* on a daily basis, including 511,408 unique wallet addresses and resulting in over 1.2 GB of data in total (see Tab. B5 in Appendix B):

- *NFT transaction data* is collected by smart contract addresses of NFT projects including trading records of different wallets, with a similar data structure to the one introduced in [56].
- *Project character data* is collected by smart contract addresses as well and contains projects' meta information (e.g., logo links and official websites) and characteristics (e.g., the creation time and floor prices of NFT projects).
- *Social media data* is collected by Twitter hashtags and reflects the popularity of NFT projects on Twitter.

For data wrangling, we mapped SQL tables into Python object classes by SQLAlchemy's ORM (Object Relationship Mapper) [71], which serves as an intermediary layer between the backend and the database to increase the efficiency of response (Fig. 2A2).

### 5.2 Data Pre-processing

We pre-processed the substantial heterogeneous data from the three sources (Sec. 5.1) into tabular numeric data.

For *NFT transaction data*, we derived the daily number of holders, sellers, and buyers in each NFT project with wallet addresses. Notably, sellers, buyers, and holders represent the three essential categories of stakeholders engaged with NFT projects during a specific period. Thus, we ensured the sum of their percentages equals one. In addition, we calculated the number of co-occurring buyers, sellers, and holders between every two NFT projects, whose percentages also add up to one. Furthermore, we applied the correlation coefficient [72] to analyze any changes in co-occurrence over time. For *social media data*, we translated all Tweets into English and categorized them based on project hashtags. To determine the prevalence of each NFT project on Twitter, we utilized a weighted sum technique that considers the daily volume of retweets, replies, likes, and quotes [51]. Moreover, we calculated the daily social media sentiment polarity for NFT projects by multiplying the sentiment polarity of each tweet, as determined by word embedding [73], with its intensity coefficient, i.e., corresponding popularity value.

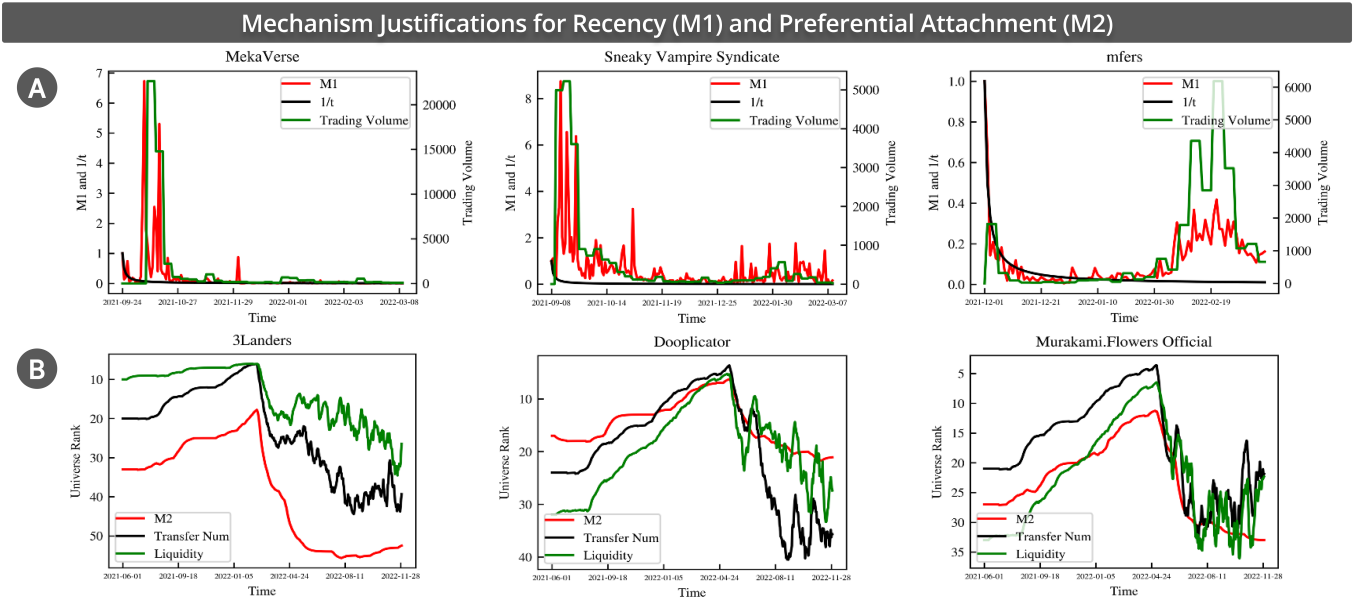

Fig. 3: (A) Three data-fitting examples compare our M1 definition (red) and the traditional method (black) against actual NFT project sales volumes (green), demonstrating that our approach is more accurate in mirroring stakeholders' perceptions of *recency* in NFT marketplaces. (B) Three examples show a positive correlation between our M2 definition (red) and the number of transfers (black) and liquidity (green) of NFT projects, underscoring our method's effectiveness and reliability.

We transformed and normalized data from three sources into numerical formats, which we then integrated into our analysis and visualization modules for deeper exploration.

### 5.3 Mechanisms Quantification

We quantify three substitution mechanisms according to their definitions and impact attributes (Tab. 2). Remarkably, the measurements for *preferential attachment* and *propensity* include two hierarchical levels: mutual and global.

#### 5.3.1 Recency

**Recency (M1)** reflects the longevity of NFT projects by combining their popularity and existing duration to gauge freshness, described as follows:

$$R_i(t) = \frac{\overline{L_i(t)}}{\Delta t}, \qquad (1)$$

Where $\overline{L_i(t)}$ indicates the average popularity of NFT projects over a certain period, and $\Delta t$ is the days since their launch. More social media discussions and newer launches within the timeframe result in a higher recency value.

*Justification*: A conventional measurement for the recency of NFT projects is solely on a project's launch dates, i.e., $1/\Delta t$. Nevertheless, through promotion on social media, prior projects can foster a resurgence of interest within their respective communities, thereby revitalizing their appeal toward potential stakeholders. Thus, NFT projects' popularity on social media hugely influences the measurement of the recency mechanism. As Fig. 3A demonstrates, the adopted M1 measurement ($E_q$ 1) aligns closely with the temporal evolution in NFT sales volume, outperforming the conventional $1/\Delta t$ method.

#### 5.3.2 Preferential Attachment

Preferential attachment, gauged by a cross-comparison of co-occurring buyers, sellers, and holders, measures the relative appeal of NFT projects. Specifically:

**Mutual Preferential Attachment** indicates the likelihood of a stakeholders transiting from project $i$ to project $j$ at time $t$, denoted as $P_{i \to j}(t)$. We align users' inflow and outflow between paired NFT projects, expressed as

$$P_{i \to j}(t) \equiv \frac{S_i(t) \cap B_j(t)}{H_i^*(t)}, \qquad (2)$$

where $S_i(t) \cap B_j(t)$ is the number of co-occurring wallet addresses concurrently buying tokens in the project $j$ ($B_j(t)$) and selling tokens in the project $i$ ($S_i(t)$). $H_i^*(t)$ is the number of wallet addresses holding the project $i$ up to $t$.

**Global Preferential Attachment (M2)** evaluates the preference level of one NFT project over others at certain times. It can be derived from the mutual preferential attachment among NFT projects and the corresponding number of holders at time $t$, as:

$$PA_i(t) \equiv \frac{\sum_{k,k \neq i} P_{k \to i}(t) H_k(t) - \sum_{j,j \neq i} P_{i \to j}(t) H_i(t)}{H_i(t)}$$
$$= \sum_{k,k \neq i} P_{k \to i}(t) \frac{H_k(t)}{H_i(t)} - \sum_{j,j \neq i} P_{i \to j}(t). \qquad (3)$$

The numerator is the expected amount of net inflow stakeholders to project $i$, normalized by the number of its holders at time $t$. A positive value indicates the number of bullish stakeholders is expected to outnumber those bearish ones on project $i$, while a negative value refers to outflows of its current holders.

Fig. 4: Illustration of quantifying the mutual substitution flows (MSFs) between NFT projects.

Fig. 5: Illustration of quantifying global impact dynamics of individual NFT projects.

*Justification*: We analyzed the correlation between M2 and both the number of transfers and the liquidity of NFT projects to assess the effectiveness of M2. Given that M2 measures the relative attractiveness of NFT projects to stakeholders, we anticipated a positive correlation with the number of transfers and liquidity. Firstly, we ranked all NFT projects based on daily data for these three metrics. Secondly, we calculated the correlation between the temporal rankings of M2 and the rankings of the other two metrics for each project. As Fig. 3B exemplifies, the temporal rankings of the three metrics show a positive correlation for most projects. Moreover, M2 displays a smoother pattern of change than the other two metrics, suggesting its reliability for gauging the preference level of NFT projects.

#### 5.3.3 Propensity

Propensity measures how similar NFT projects are by comparing their high-dimensional, inherent characteristics simultaneously. We calculate mutual and global propensity in four steps: We first utilize MiniRocket [74] to transform and classify the multivariate development sequence of NFT projects into grouped features by random convolutional kernels. Then, we provide two techniques, i.e., K-means and Gaussian mixture models, to cluster similar NFT projects with multiple impact attributes selected by the users for a specific duration (Fig. 2B2). Subsequently, we calculate the **Mutual Propensity**, i.e., $\lambda_{ij}$ in $E_q$ 5, between every two NFT projects by projecting them onto a coordinate system to detect their cosine similarity [75]. Finally, to estimate the **Global Propensity (M3)**, i.e., $\Lambda_i(t)$, we create a matrix to map the value of mutual propensity for every paired NFT project as vectors. We thus re-rank the mutual propensity distribution, yielding the estimation of the global propensity of individual NFT projects.

*Justification*: We employed feature coordinates to define M3. Algorithm 1 uses the market-cap weights and clustered feature coordinates of projects as inputs. Through mean square error optimization, it determines the market-wide feature coordinates. The cosine similarity between a project's feature vector and the market universe vector measures how much a project's features align with the entire market, defining the *Global Propensity* concept. This calculation succinctly illustrates the similarity of individual projects to the broader market, as assessed by domain experts examining the clustering results.

### 5.4 Simulating Substitutive Systems

We simulate a node-link graph for substitutive systems of NFT transactions by integrating mechanism values and stakeholders' flows between NFT projects.

#### 5.4.1 Mutual Substitution Flow (MSF)

**Mutual Substitution Flow (MSF)** denotes the direction and intensity of migrated stakeholders' inflow and outflow between paired NFT projects (Tab. 1). We adapt the evaluation of users' change rate among products in substitutive systems defined by Jin *et al.* [16] and propose $E_q$ 4 to detect the MSF between paired NFT projects:

$$F_{ij}(t) \equiv P_{i \to j}(t) \cdot H_i(t) - P_{j \to i}(t) \cdot H_j(t). \quad (4)$$

Notably, the measurement of MSF is based on calculating mutual preferential attachment values. It measures the net stakeholder flow from one NFT project to another (Fig. 4). Respectively, the minuend in $E_q$ 4 is the expected number of stakeholders who transit from project $i$ to project $j$ at time $t$, and the subtrahend evaluates the opposite flows.

#### 5.4.2 Mutual Substitution Rate (MSR)

**Mutual Substitution Rate (MSR)** captures the extent to which a specific NFT project is more likely to substitute for some counterparts than others. MSR is an essential component to gauge the impact dynamics of NFT projects

---

**Algorithm 1** Global Propensity

**Input:** Feature Data $\mathbf{v} = \{\mathbf{v}_i\}_{i=1}^{N}$ & Market cap weights $\mathbf{w} = \{w_i\}_{i=1}^{N}$

**Output:** Global propensity (M3) $\Lambda = \{\Lambda_i\}_{i=1}^{N}$

Compute market vector: $\mathbf{M} \leftarrow \mathrm{argmin}_\mathbf{M} \sum_{i=1}^{N} w_i \|\mathbf{v}_i - \mathbf{M}\|^2$

Loss function $L(\mathbf{M}) := \sum_{i=1}^{N} w_i \|\mathbf{v}_i - \mathbf{M}\|^2$

$L'(\mathbf{M}) = -2 \sum_{i=1}^{N} w_i (\mathbf{v}_i - \mathbf{M}) = 0$

$\mathbf{M} := \sum_{i=1}^{N} w_i \mathbf{v}_i / \sum_{i=1}^{N} w_i = \mathrm{argmin}_\mathbf{M} L(\mathbf{M})$

**for** $i \leftarrow 1$ **to** $N$ **do**
    $v_{i,sqr} \leftarrow 0$, $M_{sqr} \leftarrow 0$, $p_i \leftarrow 0$
    **for** $j \leftarrow 1$ **to** $K$ **do**
        $v_{i,sqr} \leftarrow v_{i,sqr} + \mathbf{v}_{i,j}^2$
        $M_{sqr} \leftarrow M_{sqr} + \mathbf{M}_j^2$
        $p_i \leftarrow p_i + \mathbf{v}_{i,j} \times \mathbf{M}_j$
    **end**
    $v_{i,norm} \leftarrow v_{i,sqr} \,\hat{}\, 0.5$
    $M_{norm} \leftarrow M_{sqr} \,\hat{}\, 0.5$
    $\Lambda_i \leftarrow p_i / (v_{i,norm} \times M_{norm})$
**end**

**return** $\Lambda \leftarrow \{\Lambda_i\}_{i=1}^{N}$

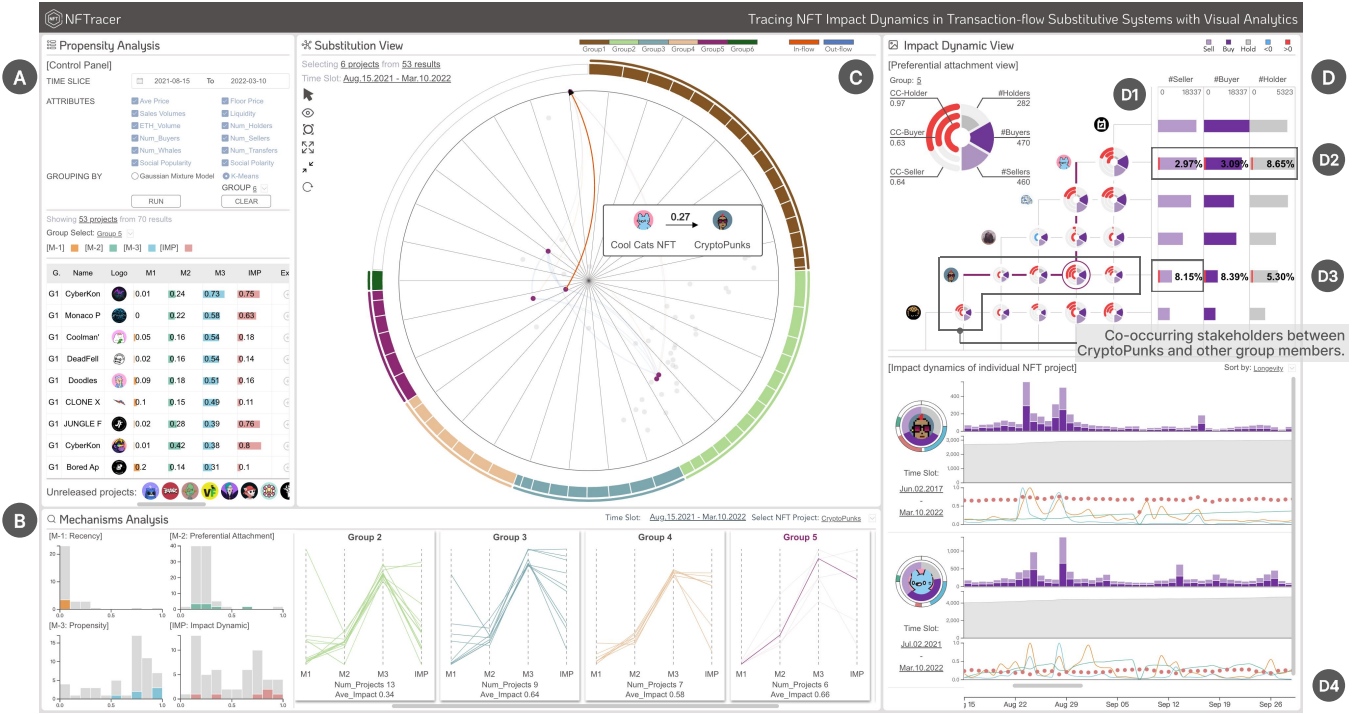

Fig. 6: *NFTracer* interface: (A) *Propensity Analysis View*, (B) *Mechanisms Analysis View*, (C) *Substitution View*, and (D) *Impact Dynamic View*. This screenshot reveals the MSF between the vintage NFT project CryptoPunks and the popular NFT project Cool Cats, their co-occurring stakeholders (D1-3), and the temporal evolution of their impact dynamics (D4).

defined in $E_q$ 6. We revised the original equation of MSR identified in the *MS model* as follows

$$\Pi_{i \to j}(t) \equiv \lambda_{ij}(t) \cdot P_{i \to j}(t) \cdot R_i(t). \quad (5)$$

$\Pi_{i \to j}(t)$ is the product of mutual propensity $\lambda_{ij}(t)$, mutual preferential attachment $P_{i \to j}(t)$, and recency $R_i(t)$ defined in Sec. 5.3. It is a unitless value to describe the likeliness of stakeholders' flow into project $j$ from project $i$.

### 5.4.3 Impact Dynamic Analysis

**Impact Dynamic** reflects the global appeal of individual NFT projects to stakeholders in substitutive systems (Fig. 5). Combining three mechanisms, we define impact dynamics by two parameters: MSR, i.e., $\Pi_{i \to j}(t)$, and the number of holders in the selected duration $H(t)$, expressed as

$$M_i(t) \equiv \sum_k \Pi_{k \to i}(t) \cdot H_k(t) - \sum_j \Pi_{i \to j}(t) \cdot H_i(t). \quad (6)$$

$M_i(t)$ estimates the project's current probability to outrun all others in terms of attracting stakeholders, i.e., impact in the whole substitutive systems. Similar to three mechanisms, $M_i(t)$ is also a continuous value that allows the user to evaluate the growth patterns of NFT projects.

## 6 VISUAL DESIGN

Our visual design aimed for a delicate equilibrium: marrying efficiency with beneficial obstructions that enhance the user experience. We streamlined the analysis process for impact dynamics, prioritizing a seamless interface. Concurrently, we crafted innovative designs such as the *Substitution View* and the *Impact Dynamic View*, each introducing a carefully measured degree of visual complexity. This intentional intricacy promotes active user engagement and cognitive processing during system interaction [76]. The ultimate goal is to strengthen long-term memory retention and deepen the understanding of the substitutive systems and underlying mechanisms governing NFT marketplaces.

*NFTracer* contains four well-coordinated views (Fig. 6). Users can start by selecting time windows and impact attributes in *Propensity Analysis View* (Fig. 6A). *NFTracer* will automatically identify and cluster NFT projects within the selected duration. Besides, users are allowed to adjust the cluster method and group numbers until it yields satisfactory results (**R1, R3**). Simultaneously, the *Mechanisms Analysis View* (Fig. 6B) will exhibit the inter-group distribution of mechanism values and impact dynamics (**R2, R4**). Then, users can highlight a group in *Substitution View* (Fig. 6C) to investigate the mutual substitution flows of NFT projects within the group (**R2, R3**). Finally, *Impact Dynamic View* (Fig. 6D) will present the preferential attachment value of NFT projects within the selected group. Furthermore, users can explore and compare the growth patterns of paired NFT projects by observing the evolution of the number of stakeholders, mechanisms, and impact dynamics (**R5, R6**).

### 6.1 Propensity Analysis View

*Propensity Analysis View* (Fig. 6A) contains two parts: the upper *Control Panel*, which supports customized analysis (**R1**), and the lower *Project List* (**R3**), which displays the tabular data of NFT projects that existed in the sliced period.

*Description*: The upper *Control Panel* function to specify users' analysis scope. Users can narrow down to a specific time window from June 2021 to November 2022 for analysis by clicking the overly open calendar from the time slice component. Along with selecting impact attributes, the clustering method, and the number of groups, users can quickly identify the data of interest and personalize the propensity values for further exploration.

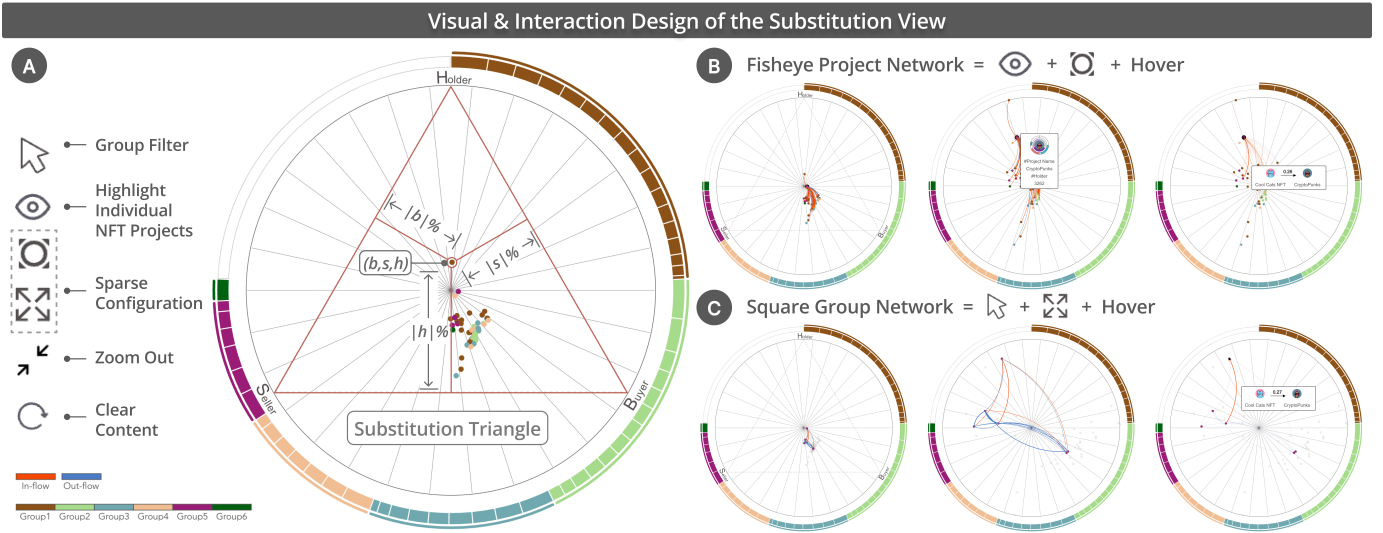

Fig. 7: Visual and interaction design of the *Substitution View*: (A) Multi-attribute-aware layout to provide fine-grained analysis of NFT projects with the toolkit and design details illustrated; (B) One example of the fisheye project network displaying the stakeholders' flow-in and out of CryptoPunks; (C) One example of the square group network exhibiting the node-link graph of stakeholders' migration among NFT projects within Group 5.

The lower *Project List* provides detailed information on all sampled NFT projects existing within the selected time window. These projects are clustered into groups and sorted by the number of projects in each group. Users can scroll through the *Project List* and expand specific projects for more detailed information. Additionally, they can focus on subsets of particular groups by the filtering interaction.

*Justification*: In addition to providing group-based tabular data on NFT projects, we explored an alternative approach to project their multi-variate attributes into a 2D space using dimension reduction techniques such as t-SNE [77]. However, the resulting visualization did not effectively reveal distinct groups. Moreover, the projected view added cognitive load for users who required detailed data to streamline their analysis workflows and focus on specific groups or group members.

### 6.2 Mechanisms Analysis View

The *Mechanisms Analysis View* (Fig. 6B) shows group-based distributions of mechanisms and impact values, facilitating inter-group comparisons for users **(R2, R4)**.

*Description*: This view consists of two juxtaposed parts. The left section displays four bar charts arranged in a quadrant layout, illustrating the value distributions for three mechanisms and the corresponding impact dynamics **(R2)**. Respectively, these mechanisms and impact dynamics are color-coded in orange, green, blue, and red. The x-axis represents the normalized values of these factors, while the y-axis indicates the number of NFT projects. Additionally, the right section features parallel coordinate plots (PCPs) that follow the group color palette. These PCPs show the values for the three mechanisms and impact dynamics across individual NFT projects, organized by group. Users can directly observe and compare the inter-group distributions of three mechanisms and impact dynamics **(R4)**. Moreover, users can highlight NFT projects of interest in *Mechanisms Analysis View* by selecting project names. Thus, users can quickly assess this project with its normalized mechanisms and impact dynamics values during the selected duration.

*Justification*: In addition to PCPs, we also experimented with radar plots to showcase the value distributions of individual NFT projects grouped accordingly. However, domain experts opined that the circular design of radar plots is not intuitive, potentially leading to a steeper learning curve. Furthermore, radar charts have scalability limitations, resulting in visual clutter, especially when dealing with a substantial number of NFT projects.

### 6.3 Substitution View

The *Substitution View* (Fig. 6C) features a *substitution wheel* with a triangular node-link graph and an outer *cluster ring*, illustrating substitutive systems. This view supports exploring the market performance of NFT projects and identifying the substitution flows between them **(R2, R3)**.

*Description*: The innermost *substitution triangle* is a node-link graph applying multi-attribute-aware techniques from the *opinion wheel* [78] for layout design. As per the requirements of domain experts, it concurrently presents stakeholder proportions across different categories to facilitate the rapid evolution of NFT projects' growth status. Meanwhile, buyers, sellers, and holders represent all relevant stakeholders of one NFT project during the selected duration, ensuring the sum of the percentage of tripartite stakeholders equals one. This rationale underpins our selection of an inscribed equilateral triangle as the projection coordinate system. Specifically, we map the status of NFT projects as vectors within the equilateral triangle with three vertices denoting the extreme status of buyers, sellers, and holders. The vectors denoting NFT projects are defined as $(b, s, h)$, where $b + s + h = 1$ (Fig. 7A). The vectors are colored based on the cluster they belong to. Directional links between vectors represent the directions and intensity of

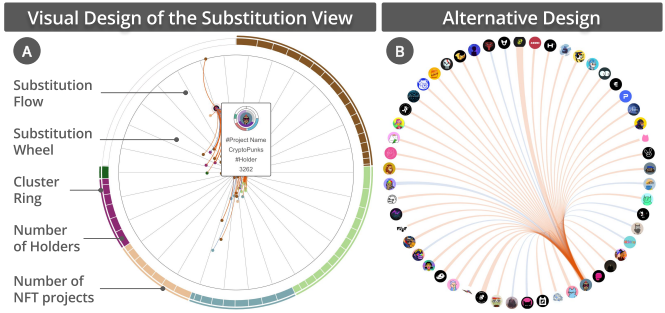

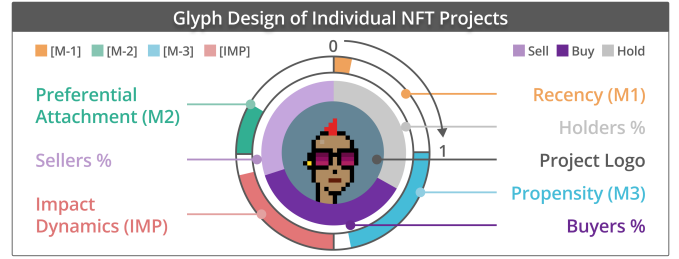

Fig. 9: The design of the *NFT status glyph* illustrates the development trend of individual NFT projects.

Fig. 8: (A) *Substitution View* presenting the node-link graph centering CryptoPunks in the fisheye sparse configuration; and (B) the alternative design utilizing a node-link graph with nodes representing NFT projects arranged in a circular layout according to their holder count.

MSF, as determined by $E_q$ 4. Therefore, users could quickly extract 1) *the substitutive systems of grouped NFT projects with ascending or descending trend visually presented* (**R2**), and 2) *NFT projects that are more likely to substitute each other* (**R3**).

The unsealed *cluster ring* comprises two layers. The narrow outermost circle represents different groups, with the length of each arc indicating the number of NFT projects within that group. These arcs are organized in a clockwise descending order, which facilitates users in assessing and exploring the outcomes of the clustering process. Synchronously, the inner, unevenly divided ring ranks NFT projects by their current number of holders. The white section of *cluster ring* denotes NFT projects that have expired or have not been launched within the duration.

This view provides four major *interactions* as follows:
**Group Filter**. By clicking the outermost *cluster ring*, the *substitution wheel* will exclusively display the selected group, its members, and in-between links. Simultaneously, the *Mechanisms Analysis View* will highlight the mechanism distribution of the corresponding group, and the *Impact Dynamic View* will be refreshed to display associated details.
**Highlight Individual NFT Projects**. Users can inspect overlay information on NFT projects by hovering over the inner circle arcs of the *cluster ring* or vectors in the node-link graph (Fig. 8A). We designed a three-layered *NFT status glyph* to encapsulate the development trend of individual NFT projects. The glyph is three-layered (Fig. 9). The top layer displays the official logos of NFT projects. The middle layer presents a pie chart featuring the ratio of buyers, sellers, and holders of NFT projects for the current time window. Lastly, the bottom layer portrays a quadratic donut chart with the length of arcs proportional to the normalized values of the three mechanisms and impact dynamics of NFT projects over the same duration. Clicking on these inner circle arcs, the *substitution wheel* will reveal all relevant substitution flows connected to the selected NFT projects, i.e., the substitutive network of individual NFT projects.
**Sparse Configuration**. To avoid visual clutter, we offer two sparse configurations for users to examine the substitution flows. One is the fisheye view [79] (Fig. 7B), and the other is square-sparse view (Fig. 7C). Both methods maintain the original relative positions of nodes denoting NFT projects.
**Clear Content**. Users can reset the *substitution wheel* and restart their analysis at any interaction stage.

*Justification*: We considered leverage chord diagram or augmented RadViz with parallel coordinates [80] to illustrate the substitutive systems. Comparatively, chord diagrams can only reveal one-dimensional attributes such as holder count in Fig. 8B, whereas using augmented RadViz with parallel coordinates requires more space. Crucially, neither of these methods efficiently showcases the status of NFT projects. As $E_B$ stated, "*The appeal of NFT projects to stakeholders is reflected in their transaction behaviors. In other words, the proportion of holders, sellers, and buyers in the stakeholder pool can indicate whether the impact of the current NFT project is in a rising or declining phase.*" As such, our multi-attribute-aware technique is deemed superior in providing comprehensive insights.

### 6.4 Impact Dynamic View

According to domain experts, studying the impact dynamics of NFT projects requires a two-dimensional comparison. This involves examining the number of co-occurring stakeholders and the correlation between their volume changes (**R2, R3**). It also involves analyzing the temporal growth pattern of individual NFT projects (**R5, R6**). To facilitate this analysis, the *Impact Dynamic View* (Fig. 6D) presents the bipartite set relations of NFT projects through the top *Preferential Attachment View*. Meanwhile, the bottom *Individual Evolution View* illustrates their temporal development.

*Description*: The upper *Preferential Attachment View* (Fig. 6 D1) demonstrates **co-occurring stakeholders** of paired projects within each cluster, alongside the correlation coefficients of three behavioral dimensions, i.e., sell, buy, and hold (**R2, R3**). To present the co-occurring stakeholders, we measured the number of recurring wallet addresses between pairs of NFT projects. We also calculated the correlation coefficients for the temporal evolution of three types of stakeholders. We then compared state-of-the-art visualizations for bipartite relationships. Ultimately, we adapted *Aggreset* [81] to display and correlate the co-occurring stakeholders across all three categories within a *set matrix*.

The *set matrix* comprises *stakeholder co-occurrence glyphs*, each split into two halves (Fig. 10A) [82]. The right half's semicircular rose diagram shows three equally angled sectors for buyers (dark violet), sellers (lilac), and holders (grey) in NFT project pairs, with radial distances indicating their proportions. The left half's concentric donut plots reveal the correlation coefficients for stakeholders' temporal evolution. Angles indicate the coefficients' absolute values, with red for positive and blue for negative. The magnitude of each glyph signifies the quantity of co-occurring stakeholders in three categories. The right half of *Preferential Attachment View* comprises three horizontally aligned histograms, showing the number of sellers, buyers, and holders of individual projects during the selected duration. The sum of stakeholders ranks the aligned histograms in three categories.

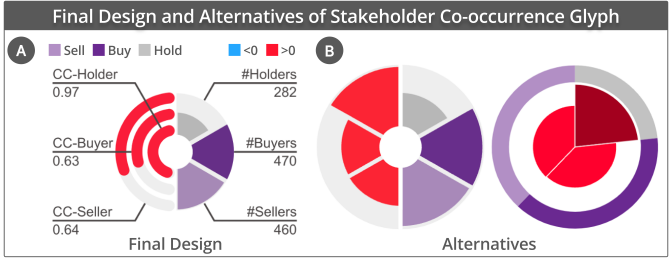

Fig. 10: The final glyph design (A) and two alternatives (B) exhibit the co-occurring stakeholders between two NFT projects and their evolution correlation coefficient values.

The *Individual Evolution View* located at the bottom displays a comprehensive comparison of NFT projects' temporal evolution (**R5, R6**). This view combines a stacked bar chart, an area chart, a line chart, and a scatter plot to deliver the growth patterns of NFT projects. It demonstrates the daily fluctuations of stakeholders in three categories, the value of mechanisms, and impact dynamics for individual NFT projects throughout the selected period. The glyphs representing individual NFT projects are also incorporated into this view. Summarizing the data is crucial when analyzing NFT market performance, considering price and popularity volatility. This representation reminds users about long-term market performance and potential rather than solely focusing on short-term trends.

This view provides two *details on demand* interactions:
**Co-occurrence Highlighting**. Via clicking *stakeholder co-occurrence glyphs* in the *set matrix*, the right histograms display short vertical lines (Fig. 6D2). These lines illustrate the proportion of co-occurring stakeholders within their respective categories in the chosen paired NFT projects (Fig. 6D3). The color coding of these lines signifies whether the correlation coefficient is positive or negative.
**Paired Comparison**. Selecting the *stakeholder co-occurrence glyphs* in the *set matrix* reveals the *Individual Evolution View*, illustrating the temporal progression of NFT projects. This view allows for ranking projects by their longevity, impact dynamics, and overall stakeholder engagement.

*Justification:* To design the *Preferential Attachment View*, we surveyed the state-of-the-art visualization techniques that depict set relationships for domain experts to compare. We considered four candidate approaches, including *OnSet* [83], *Radial Sets* [84], *AggreSet* [81], and *UpSet* [85]. Domain experts posited all as capable of analyzing co-occurring stakeholders in paired NFT projects. However, *OnSet* necessitates re-defining matrix encoding and is not intuitive for displaying three categories of stakeholders. On the other hand, *UpSet* and *Radial Set* require larger display space, that may be considered overqualified as they excel in one-to-many or many-to-many comparison scenarios. Thus, we modified the *Aggreset* due to its intuitive display of co-occurring stakeholders between similar projects, enabling rapid bipartite comparisons. Additionally, we streamlined project filtering through a cross-view interaction (see details in Sec. 6.3), providing simplicity and space efficiency.

For the *Individual Evolution View*, we initially intended to use dual y-axes to consolidate data from three types of stakeholders. However, we recognized the risks of user confusion and challenges with direct comparisons. To mitigate these issues, we created separate area charts for the larger dataset, the number of holders. Furthermore, we strategically used scatter plots and line charts to distinguish between *impact dynamics* and *mechanisms*. We combined both visual methods, considering that impact dynamics act as dependent variables for the three mechanisms and serve as key indicators for users evaluating NFT projects' market appeal. The fusion of line charts and scatter plots enables domain experts to compare metrics more effectively.

## 7 CASE STUDIES

This section presents two case studies based on real NFT transaction data with domain experts $E_A$ and $E_C$. They freely explored *NFTracer* and gained insights into the growth patterns and substitute relationships among NFT projects. Each case study lasted for around one hour.

### 7.1 Case 1: Characterizing "(Un)healthy" NFT Projects

Identifying the growth patterns differentiates "healthy" NFT projects, with longer lifespans and better performance, from "unhealthy" ones showing the reverse (**R1, R4, R5**).

**Insight 1: Stable and gradual change in mechanisms is the key to maintaining a "healthy" trend for NFT projects** (Fig. 11A). $E_A$ characterized the growth patterns of "healthy" NFT projects by exploring the distribution and development of mechanisms. $E_A$ started by slicing the time window from June to September 2022 and selected all attributes to cluster similar NFT projects into seven groups (**R1**). Then he noticed from the *Propensity Analysis View* that some "healthy" NFT projects had been clustered into Group Three, including Dooplicator, Doodles, 0N1 Force, and Bored Ape Kennel Club. Next, he investigated the *Mehchanisms Analysis View* and found their propensity (M3) values to be relatively low (**R4**). Since M3 measures how similar NFT projects are, he interpreted that, "*lower M3 values could indicate unique growth patterns with a higher possibility of survival.*" As such, $E_A$ explored the *Impact Dynamic View* to analyze the evolution of individual projects (**R5**). He observed that the value of recency (M1) exhibited cyclic growth since the launch of projects. This pattern echoes his domain knowledge about how brokers periodically enhanced the popularity of NFT projects on social media to refresh stakeholders' impressions. Moreover, he found that preferential attachment (M2) values decreased evenly in well-performing projects. Particularly, $E_A$ pinpointed one atypical "healthy" project, CryptoPunks, whose M2 value consistently remained high. He stated, "*CryptoPunks occupies a distinctive position as an 'antique' in the NFT community, which reduces stakeholders' willingness to sell.*"

**Insight 2: Dropping stakeholders' attention and interest can lead to "unhealthy" churns** (Fig. 11B). With similar analysis pipelines, $E_A$ also identified growth patterns of "unhealthy" NFT projects, e.g., MekaVerse, Monaco Planet Yacht, and SupDucks. He concluded from the *Impact Dynamic View* that M1 decreases rapidly after the initial launch in these projects, indicating a deficiency in communication between the project launcher and stakeholders (**R5**). Furthermore, their M2 value tends to be relatively low, presenting a failure in maintaining attractiveness to stakeholders (**R4**). Particularly, $E_A$ observed the substitution flows in the *Substitution View* and noted an escaping trend of stakeholders owning aforesaid projects (**R1**), which denotes lower user loyalty with many stakeholders migrating from these projects to more promising ones. $E_A$ explained, "*Stakeholders can leave for multiple reasons, such as launchers' inability to fulfill the promised road map, rug and pull strategies, and no appropriate incentives for community members.*"

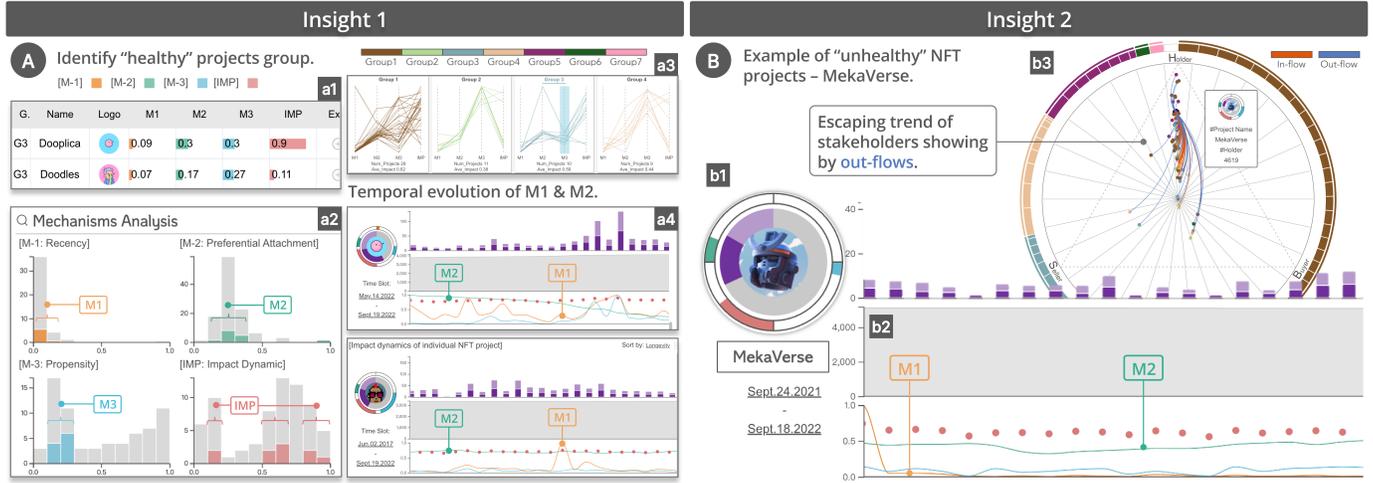

Fig. 11: Illustration of case 1. (A) $E_A$ sequentially utilized the *Propensity Analysis View* (a1), *Mechanisms Analysis View* (a2 & a3), and *Impact Dynamic View* (a4) to identify and characterize "healthy" NFT projects. (B) He detected features of "unhealthy" NFT projects from the *Impact Dynamic View* (b1 & b2) and the *Substitution View* (b3).

### 7.2 Case 2: Compare Competitive NFT Projects

Stakeholders can refine their investment portfolios by identifying and comparing mutually substitutable NFT projects (**R2, R3, R6**), yielding two key insights from $E_C$.

**Insight 1: NFT projects featuring a higher degree of similarity are more likely to substitute each other** (Fig. 12A). $E_C$ verified his hypothesis that similar projects were more likely to mutually substitute by comparing the values of propensity and substitution flows. He started by observing the group-based sub-dataset from April to July 2022 with all attributes selected in the *Propensity Analysis View*. While inspecting the *Mechanisms Analysis View*, he noticed that NFT projects in Group Four tended to possess higher propensity values (**R2**). $E_C$ then used the *Substitution View* to inspect substitution flows within this group (**R3**). Next, he randomly highlighted individual NFT projects belonging to this group and examined their relevant node-link graphs. By comparing the thickness of substitution flows and hovering over specific flow values, $E_C$ found that substitution flows between group members with higher propensity values tended to be thicker than nonmembers, implying a higher possibility of substitution. Hence, he concluded, "*While stakeholders move freely among various NFT projects, those sharing similarities tend to have more migrated stakeholders between them.*"

**Insight 2: Competitive NFT projects akin to mechanism distributions are negatively correlated in impact dynamics values** (Fig. 12B). $E_C$ compared co-occurring stakeholders and the impact dynamics evolution among similar projects to analyze competitive ones. Upon filtering Group Four, he noticed in the *Impact Dynamic View* that World of Women (WoW) and World of Women Galaxy (WoWG) shared the most co-occurring stakeholders, with the number of sellers and buyers negatively correlated (**R3**). Thus, he commented, "*The buyers and sellers of these two NFT projects share a reciprocal relationship. However, their holder base is positively correlated, suggesting that they have overlapping target audiences and are in competition.*" Furthermore, $E_C$ selected these two projects in *Mechanisms Analysis View* and found that despite having comparable mechanism distributions, notable differences existed in their impact dynamics values (**R2**). He then analyzed the *Individual Evolution View* of paired NFT projects, detecting that WoWG had lower M2 values and impact dynamics than WoW (**R6**). Moving to the *Substitution View*, he observed a notable stakeholder migration from WoWG to WoW in the substitution flow as expected.

## 8 USER STUDY

This section introduces the procedure and results of evaluating *NFTracer* with 13 NFT stakeholders.

### 8.1 Participants and Procedure

We recruited 13 stakeholders (S1–13: M = 6, F = 7) from NFT communities, such as WeChat groups and Twitter spaces, with diverse backgrounds (please refer to Tab. C6 in Appendix C). These stakeholders have various market roles, including investors, critics, scholars, and collectors, aged between 24-32 (Avg = 27.62, SD = 4.24). Moreover, their experiences with NFT transactions range from 0.4 to 2.5 years (Avg = 1.14, SD = 0.43).

In this study, we first collect demographic information and consent forms from the participants. Then, we introduce them to the key concepts relevant to the substitutive systems of NFT transactions, followed by the design and workflow of *NFTracer*. Specifically, we demonstrated our system's workflow using a pre-trained dataset from August 15, 2021, to March 10, 2022. We started by introducing the control panel within the *Propensity Analysis View*, then applied the K-means method with all impact attributes selected and set six clusters for NFT projects. We detailed the three mechanisms and impact dynamics, along with their visual encoding, through the *Mechanisms Analysis View* for the participants. We also showcased the visual encoding and interactive features of the *Substitution View* and *Impact Dynamic View*. During the 20-minute presentation, participants were encouraged to freely interrupt with questions.

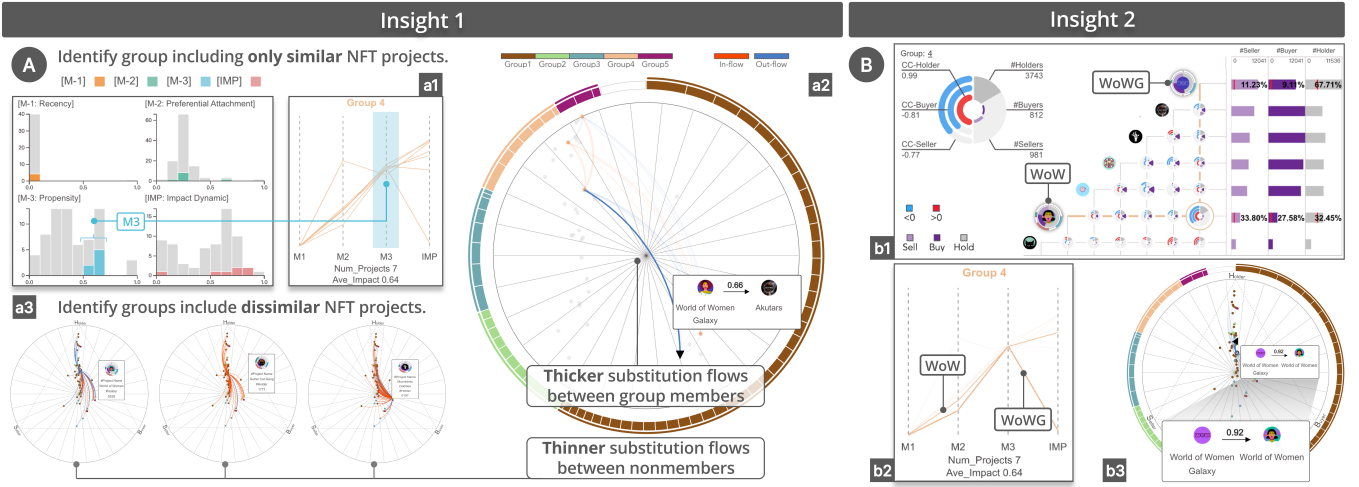

Fig. 12: Illustration of case 2. (A) $E_C$ sequentially explored the *Mechanisms Analysis View* (a1), and *Substitution View* (a2 & a3) to compare networks of similar NFT projects with dissimilar ones. (B) He compared two competitive NFT projects by sequentially investigating the *Impact Dynamic View* (b1), *Mechanisms Analysis View* (b2), and *Substitution View* (b3).

Afterward, we let participants freely explore *NFTracer* and familiarize themselves with its functions (10 mins). We continually observed their behavior during this process. We addressed the participants' inquiries concerning the system's visual and interaction design until they were ready for three user study tasks, which we then timed as they completed. Later, we conducted semi-structured interviews to gather their feedback and suggestions on using the system (see responses in Tab. C8). Finally, we ask the participants to evaluate *NFTracer* using a seven-point Likert scale (see questions in Tab. C7). The entire user study is conducted and recorded on Zoom, taking about 75 minutes.

### 8.2 Task Completion

Participants were invited to complete three analysis tasks derived from six design requirements (see Sec. 3.4) following the think-aloud protocol [86], which are: finding similar NFT projects that have a higher mutual substitution possibility ($T_a$); identifying and compare inter-group mechanism distributions ($T_b$); and selecting and comparing the time-varying growth patterns of individual NFT projects ($T_c$). We informed the participants that there was no single correct approach or answer for completing each task in advance. $T_a$ takes an average of 8.04 minutes (SD = 2.90) as this task requires a comprehensive analysis, and stakeholders have limited trust in automatic clustering outcomes for NFT projects. They prefer to leverage the *Mechanisms Analysis View* or *Impact Dynamic View* to identify similar projects. Furthermore, some participants found detecting mutually substitutive projects time-consuming, especially when sub-stitution flows were dense. In contrast, $T_b$ and $T_c$ took less time to complete, averaging less than five minutes, as participants can easily identify the relevant views for intergroup or individual project comparisons.

### 8.3 Measures and Results

We used semi-structured interviews and a seven-point Likert scale [56] to assess the *informativeness, effectiveness, design,* and *usability* of *NFTracer*, with results shown in Fig. 13.

**Informativeness (Q1-3)**. Participants generally reported that *NFTracer* provided sufficient information to gain insights into the substitutive systems for NFT transactions as desired. They appreciated the *details on demand* interactions in the *Substitution View* and *Impact Dynamic View*, which helped investigate the substitution flow and co-occurring stakeholders between NFT projects. Some suggested including more information about defining and quantifying mechanisms to shorten their learning time and facilitate the usage of *Mechanisms Analysis View*.

**Effectiveness (Q4-7)**. Overall, participants found the system effective. They mentioned that quantified mechanisms and impact dynamics facilitated their hierarchical understanding and comparison of the substitutive systems of NFT transactions. Particularly, they mentioned that the *set matrix* and *stakeholder co-occurrence glyph* were impressive and helped deepen their understanding of the *Preferential Attachment* mechanism. S4 said, *"The glyph design was harmonious and effective, which increased my interest in exploration."* This feedback aligns with the findings of Hullman *et al.* [76], suggesting that well-designed visual difficulties can enhance users' perception and understanding.

**Visual Design and Interaction (Q8-12)**. In general, participants acknowledged all views as functional and intuitive. Most highlighted the *Impact Dynamic View* as appropriate and straightforward for them to compare the substitutive relationships between NFT projects. Some commented on the *Substitution View* as novel and interesting to explore. However, they suggested slightly extending the pixels of the node-link graph within the *substitution wheel* to smooth relevant interactions. For example, S2 opined that *"I think*

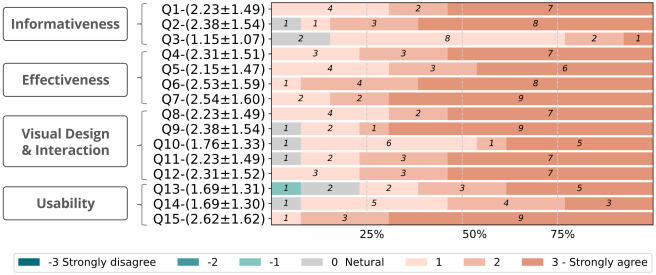

Fig. 13: Seven-point Likert scale results: The middle column shows Mean ± SD adjacent to the bar chart.

*it would be better to expand the node-link graph in Substitution View a little bit more, and thus, I can compare the amount of migrated stakeholders more quickly."*

**Usability (Q13-15)**. Most participants found *NFTracer* useful for NFT impact dynamics analysis. Specifically, collectors and researchers (S2, S10, S12) noted that our system, particularly *Mechanisms Analysis View*, assisted them in quickly identifying and filtering NFT projects of interest. For instance, S2 appreciated the group-based PCPs for intuitive inter-group comparisons, which streamlined his analytical framework. Meanwhile, investors (S6, S11) highlighted *NFTracer*'s capability for tracking macro market trends by detecting stakeholder migration. S6 was pleased with the system's overall workflow and has already endorsed it to his peers. Furthermore, he proposed integrating Etherscan and Twitter APIs for real-time data or connecting his company's database for future use in his routine analytics.

# 9 DISCUSSION

**Lessons Learned**. We learned lessons from collaboration with domain experts when developing *NFTracer*. First, a domain-driven design is beneficial for clarifying the research scope and developing the model. We discovered that domain experts presupposed influencing factors for evaluating NFT projects, whereas needing help quantifying them. Visually presenting potential models, discussing with domain experts, and gathering stakeholders' responses can identify the mechanisms and impact attributes needed for simulating the model. Second, flexible interactions matter when the system serves users from diverse backgrounds. In our case, individuals prefer substitutive periods and impact attributes to simulate the substitutive systems of NFT transactions, which necessitate flexible interactions. Third, we find that proficiency in visualization literacy can influence users' perceptions of usability. Despite providing consistent demonstrations of the system to all participants, individuals with expertise in visualization and mathematics tended to provide more positive usability evaluations. Participants in humanities, social sciences, and arts found learning the *NFTracer* more challenging than those from STEM fields. For instance, these participants (S7, S8, S13) took longer to adjust to the visual analytics approach. This observation might suggest a higher cognitive and learning demand for individuals less accustomed to complex data dashboards when transitioning to visual analytics. However, these correlations are not definitive, warranting further research to explore the impact of their research practices on visual analytics proficiency. Last but not least, model interpretation facilitates stakeholder exploration. We found that stakeholders lack confidence in the algorithm's automatically generated results when they have a limited understanding of its underlying process. This skepticism is further compounded if the independent variables employed by the algorithm do not align with their personal experiences. Thus, balancing efficiency and explainability by providing fine-grained visual presentations for underlying mechanisms is valuable.

**Applicability**. Our *NFTracer* system supports both one-off analysis and continuous monitoring of NFT marketplaces. Some participants (S1, S5, S6, S11) engaged in multiple analyses over different time windows when completing $T_c$ in the user study, demonstrating *NFTracer*'s utility beyond one-off analyses, catering to both sporadic and regular monitoring needs in the NFT market. Furthermore, *NFTracer* transcends its role as merely an analysis and monitoring tool within the NFT industries. It enhances the explainability of the *MS Model*, a crucial method that enables researchers to adeptly navigate the rapidly evolving global techno-economic landscape [34]. This enhancement in model transparency not only fosters a deeper understanding of impact dynamics but also empowers users to customize the *MS Model* for their unique areas of study or applications. As a result, enhancing the interpretability of the *MS Model* extends the long-term applicability of *NFTracer* to a broader spectrum of practical and research scenarios.

**Significance and Generalizability**. Impact dynamics analysis is crucial for understanding innovation processes [34]. The late 20th-century digital revolution not only sparked swift changes in concepts, products, and technologies but also necessitated producers monitoring the turnover of these fast-paced entities to stay current with rapidly shifting market trends [87]. Consequently, *NFTracer* has targeted NFT marketplaces, given the unique and fast-evolving nature of NFT data. However, the system also offers new opportunities for researchers and institutions. It can be tailored to study competition across various creative industries, uncovering paradigm shifts in academia or consumer trends in retail. Moreover, the *MS Model* is highly adaptable, supporting the combination of impact attributes and the integration of new variables for a wide range of applications to meet diverse research objectives. For instance, the model can be reconfigured for new domains by altering attributes, such as replacing price with citation counts in scholarly research.

**Limitations and Future Work**. *NFTracer* is limited in two aspects. First, *NFTracer* could be improved by more fine-grained modeling. We currently consider three major stakeholder categories, i.e., buyers, holders, and sellers, excluding one transitional status, which is "listing" tokens from NFT collections to sell. Moreover, the relevant market information on NFT transactions is limited, yet involves more potential attributes, e.g., cryptocurrency market volatility, which could be beneficial for analysis. Thus, we will further broaden attributes in our data set to improve the accuracy of our model. Second, the *MS Model* cannot explicitly detect the feature importance of attributes and mechanisms to NFT impact dynamics. While stakeholders can infer interrelationships between attributes, mechanisms, and growth patterns by inspecting temporal evolutions, they cannot accurately compare their significance. Future work can leverage variable selection techniques to further interpret this model and predict the growth patterns of NFT projects accordingly. We are also integrating *NFTracer* with real-time data, which is essential for collectors and investors who need to monitor market trends and transactions frequently. With this feature, our system will sustain long-term user engagement, and we will evaluate the performance and reception of *NFTracer* over time. Furthermore, we will collect user interaction logs with the consent of domain experts to sample and analyze feedback on *NFTracer*'s usage in real workplaces.

## 10 CONCLUSIONS

In this work, we propose an analysis framework for stakeholders to detect the impact dynamics of NFT projects. We first extracted analysis criteria, i.e., impact attributes and mechanisms, through a formative study with domain experts and stakeholders. Then we constructed the *Minimal Substitution (MS) Model* along with intuitive visualizations to simulate the substitutive systems of NFT transactions as node-link graphs. Particularly, we present two sparse configuration techniques to resolve the visual clutter issue. Accordingly, we developed *NFTracer*, an interactive VA system for stakeholders to hierarchically explore the substitutive systems of NFT transactions and identify growth patterns of NFT projects. Finally, we conducted two case studies and one user study with target users to demonstrate the informativeness, effectiveness, visual design, and usability of *NFTracer*. In the future, we plan to include more relevant attributes and leverage variable selection techniques to improve the accuracy of the *MS Model* and help stakeholders identify the most significant factors.

### ACKNOWLEDGMENTS

The authors would like to thank Songyu Wang, Dr.U (Twitter: @dr_uzuz), Lizhong Chen, Lifang Zheng, Laixin Xie, Xiyuan Wang, and Shauna Dalton for their help and the Crypto-Fintech Lab of HKUST and China Capital International Cooperation (International) for their collaboration. The authors would like to thank the anonymous reviewers for their valuable comments. This work is partially supported by the Hong Kong Research Grant Council (GRF16210722).

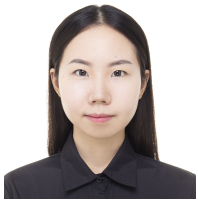

**Yifan Cao** is a Ph.D. candidate in the Individualized Interdisciplinary Program (CMA) at the Hong Kong University of Science and Technology. She obtained her Bachelor's Degree in Chinese Literature and Linguistics from Chu Kochen Honors College at Zhejiang University and completed her Global Communication M.A. at the CUHK. Her research interests include social computing and visual analytics.

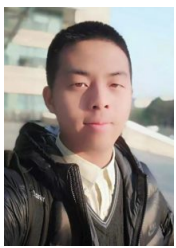

**Qing Shi** was a research assistant at the Computational Media and Arts (CMA) Thrust of the Hong Kong University of Science and Technology (Guangzhou). His recent research interests include visualization and visual analytics, XAI, and HCI.

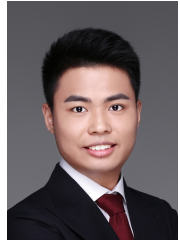

**Lue Shen** is currently a Ph.D. candidate in Mathematics at the Hong Kong University of Science and Technology. He is also working as a system developer at China Capital International Cooperation (International). He received BS in Computing Mathematics from the City University of Hong Kong and MS in Financial Mathematics from HKUST. His research interests include natural language programming, distributed ledger technology, and financial mathematics.

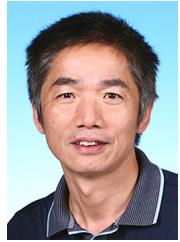

**Kani Chen** holds a joint appointment as a Professor in both the Department of Mathematics and the Department of Industrial Engineering and Decision Analytics at the Hong Kong University of Science and Technology (HKUST). He also serves as the Program Director of the MSc in Financial Mathematics and Co-Director of the Risk Management and Business Intelligence Program. Prof. Chen received his BSc and MSc from Peking University and earned his PhD in Statistics from Columbia University in the City of New York. His current research interests include survival analysis, sequential analysis, boot-strapping, empirical process, stochastic modeling, missing data and EM algorithm.

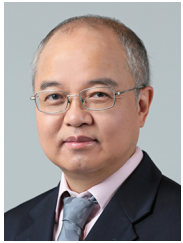

**Yang Wang** currently serves as the Vice-President for Institutional Advancement at the Hong Kong University of Science and Technology (HKUST). He is a Chair Professor in the Department of Mathematics and Industrial Engineering and Decision Analytics in HKUST. Additionally, he is the Director of Big Data for the Bio Intelligence Laboratory and Associate Director of the Big Data Institute at HKUST. Prof. Wang received his Bachelor's degree in Mathematics from the University of Science and Technology of China and earned his Ph.D. in Mathematics from Harvard University. Prof. Wang has a wide range of research interests, including data analysis using machine learning, fractal geometry, signal processing, wavelets and analysis, and wavelets and frames. He has published over 100 research journal papers in both pure and interdisciplinary mathematics.

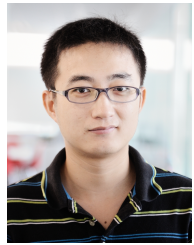

**Wei Zeng** is an assistant professor at the Hong Kong University of Science and Technology (Guangzhou). He received his Ph.D. in computer science from Nanyang Technological University in 2015. He received ICIV'15 and VINCI'19 Best Paper Award and ChinaVis'21 Best Paper Honorable Mention Award. He serves as Program Chair for VINCI'23, a program committee for venues including IEEE VIS, EuroVis STARs, and ChinaVis. His recent research interests include visualization and visual analytics, computer graphics, AR/VR, and HCI.

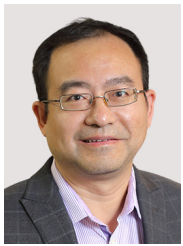

**Huamin Qu** is currently the dean of the Academy of Interdisciplinary Studies (AIS), the head of the Division of Emerging Interdisciplinary Areas (EMIA), and a chair professor in the Department of Computer Science and Engineering (CSE) at the Hong Kong University of Science and Technology (HKUST), and was the acting head of the Computational Media and Arts Thrust Areas (CMA) at HKUST(GZ). He obtained a BS in Mathematics from Xi'an Jiaotong University, China, an MS, and a PhD in Computer Science from Stony Brook University. His main research interests are visualization and human-computer interaction, focusing on urban informatics, social network analysis, E-learning, text visualization, and explainable artificial intelligence (XAI).

# APPENDIX A

## A.1 Materials of the candidate models proposed for domain experts' discussions in the focus group

TABLE A1: Details of the seven candidate models, including the basic formulation, key mechanisms or parameters, application examples, and references. All variables $I(t)$ appearing in the equation below represent the size, population, or connected edges at time $t$, indicating the impact of specific products, services, viruses, or species. Additionally, I(0) indicates the initial impact.

| ID | Model Name | Model Formulation | Reference |
|----|-----------|-------------------|-----------|
| C1 | Bass Diffusion Model | $\frac{dI}{dt} = (p + q\frac{I(t)}{m})(m - I(t))$ | [28], [36], [37] |
| | $p$ (Innovation Factor): The likelihood of individual adoption due to personal interest or discovery. | | |
| | $q$ (Imitation Factor): The likelihood of individual adoption due to social influence or imitation. | | |
| | $m$ (Potential Market size): The maximum number of eventual adopters, setting an upper limit on cumulative adopters, I(t). | | |
| | Application examples: Detect the process of how new products, technologies, or ideas spread through a population over time. | | |
| C2 | Lotka-Volterra Competition (LVC) Model | $\frac{dx}{dt} = r_1 x \left(1 - \frac{x + \alpha y}{K_1}\right), \frac{dy}{dt} = r_2 y \left(1 - \frac{y + \beta x}{K_2}\right)$ | [88] |
| | $r_1, r_2$: Intrinsic growth rates for species 1 and 2, respectively, determining each species' isolated population growth rate. | | |
| | $K_1, K_2$: Carrying capacities for species 1 and 2, respectively, representing each species' maximum sustainable population without competition. | | |
| | $\alpha, \beta$: Competition coefficients for species 1 and 2, respectively, measuring one species' impact on the other's growth due to limited resources. | | |
| | Application examples: Describe the dynamics of two species competing for the same limited resources. | | |
| C3 | Susceptible-Infected-Recovered (SIR) Model | $\frac{dS}{dt} = -\beta \frac{SA}{N}, \frac{dA}{dt} = \beta \frac{SA}{N} - \gamma A, \frac{dR}{dt} = \gamma A.\ I(t) = A(t) + R(t)$ | [30], [31], [60], [61] |
| | $S$: Susceptible (potential users), $A$: Infectious (current users), $R$: Recovered (previous users). | | |
| | $\beta$: The transmission rate, defining the likelihood and frequency of disease transmission from infected individuals to susceptible individuals. | | |
| | $\gamma$: The recovery rate, representing the rate at which infected individuals recover from the disease and become immune . | | |
| | Application examples: Track the progression of an infectious disease by estimating how individuals move between these compartments over time. | | |
| C4 | Evolving Network Model | $\Pi_i = \frac{I_i \eta_i}{\sum I_j \eta_j}, \ \Pi_i \propto \eta_i I_i(t)$ | [25], [89] |
| | $\eta_i$: Node $i$'s fitness, an intrinsic property influencing the likelihood of gaining new connections. | | |
| | $\sum I_j \eta_j$: Sum of the degrees of all nodes in the network. | | |
| | Application examples: Capture the preferential attachment of nodes in the network at a certain time, which indicates the dynamics of social networks, biological systems, and technological networks. | | |
| C5 | Minimal Substitution (MS) Model | $\Pi_{i \to j}(t) = \lambda_{ij} N_j(t) \frac{1}{t_j}, \ \eta_i \equiv \sum_k \lambda_{k \to i} N_k, \ I_i(t_i) = h_i t_i^{\eta_i}$ | [16], [33] |
| | $\lambda$: The propensity mechanism which embodies the innate likelihood of substitutions between two products, $i$ and $j$. | | |
| | $N_j(t)$: The preferential attachment mechanism implies that individuals tend to select products of greater popularity. | | |
| | $\frac{1}{t_j}$: The recency mechanism that represents the degree to which a product's novelty diminishes over time. | | |
| | $\eta_i$: The fitness parameter that measures the the total propensity for users to switch from all other products to $i$. | | |
| | $h_i$: The anticipation parameter that captures users' initial excitement for the product $i$. | | |
| | Application examples: Detect the substitution rate between two products and also the process of how new products, technologies, or ideas spread through a population over time. | | |
| C6 | Product Competition Models | $I(t) = \prod_{s=1}^{t} (1 + r_s X_s), I(0) \approx \prod_{s=1}^{t} e^{r_s X_s}, I(0) = e^{\sum_{s=1}^{t} r_s X_s I(0)}$ | [26], [27] |
| | $r_s$: The decay parameter, characterizing the rate at which the novelty of information declines within a specified time interval. | | |
| | $X_s$: This parameter signifies positive, independent, and identically distributed random variables. | | |
| | Application examples: Illustrate the popularity dynamics of information in social networks over time during the competition for collective attention. | | |
| C7 | Gompertz Model | $I(t) = A * \exp(-B * \exp(-Ct))$ | [22], [90], [91] |
| | $A$: The upper asymptote or carrying capacity, representing the maximum attainable size or population. | | |
| | $B$: A positive constant that is related to the initial size or population. | | |
| | $C$: The growth rate constant, determining the steepness of the curve. | | |
| | Application examples: Characterize the growth patterns of social networks, e.g., the expansion of populations, tumor development, and biological phenomena. | | |

*Specifically, C1, C4, C5, C7 are generalized models that could be applied to various domains; C2 is primarily applied in ecology refers to the dynamics of two species competing for limited resources; C3 is mainly applied in epidemiology to study the spread of infectious diseases in a population; C6 is created in communication domains to detect collective attention among different pieces of information on social networks.

## A.2 Model fitting results of *Bass Diffusion Model* (*C1*), *Minimal Substitution (MS) Model* (*C5*), and *Gompertz Model* (*C7*) out of the seven candidates during the focus group discussion.

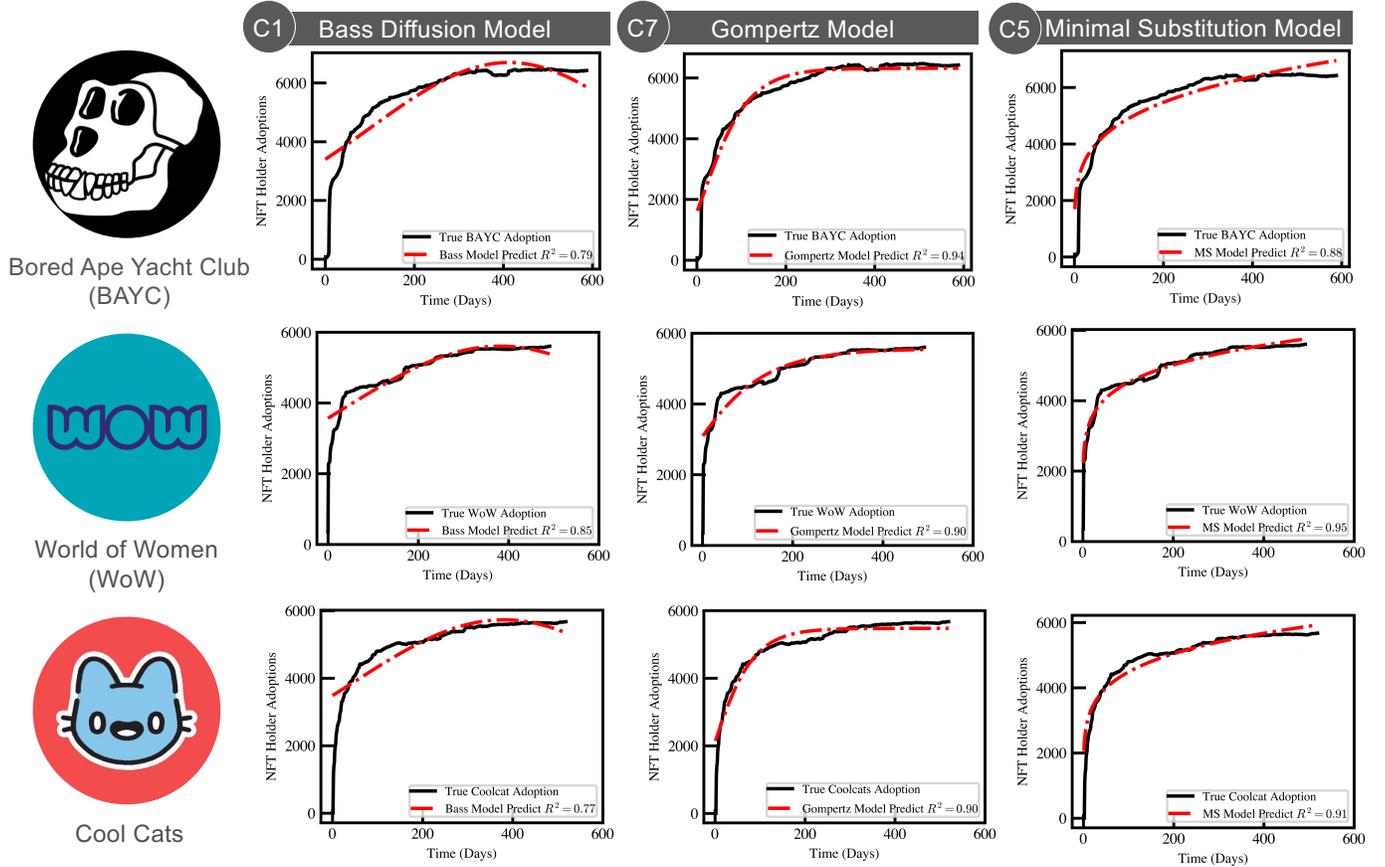

Fig. A1: Comparisons between ground-truth holders' evolution (solid black line) and model fit (red dashed line) of three NFT projects launched on different dates with different market ranks (see Table B4), i.e., Bored Ape Yacht Club (BAYC), World of Women (WoW), and Cool Cats with selected models (*C1*, *C5*, and *C7*). Based on the model fitting results measured by $R^2$, as well as the involved mechanisms and parameters, domain experts ultimately selected the *Minimal Substitution Model (MS)* for our research case.

TABLE A2: Respective parameter details of the aforementioned model fitting results (*C1*, *C5*, and *C7*) of three NFT projects.

| | Model Names | | |
|---|---|---|---|
| Project Names | C1: Bass Diffusion Model | C5: Minimal Substitution Model | C7: Gompertz Model |
| Bored Ape Yacht Club (BAYC) | $m = 5400999.719294338$ | $h_i = 1.69208619\text{e+}03$ | $A = 6.31920952\text{e+}03$ |
| | $p = 0.0006274850921316063$ | $\eta_i = 2.21661584\text{e-}01$ | $B = 1.38015309\text{e+}00$ |
| | $q = 0.0035935532284403974$ | | $C = 1.73100152\text{e-}02$ |
| World of Women (WoW) | $m = 4912992.63549968$ | $h_i = 2.23442649\text{e+}03$ | $A = 5.56073153\text{e+}03$ |
| | $p = 0.0007243552683110068$ | $\eta_i = 1.52586865\text{e-}01$ | $B = 5.91835348\text{e-}01$ |
| | $q = 0.00293585397760822$ | | $C = 1.01545931\text{e-}02$ |
| Cool Cats | $m = 4826322.792011898$ | $h_i = 2.05787671\text{e+}03$ | $A = 5.47589969\text{e+}03$ |
| | $p = 0.0007204043831322507$ | $\eta_i = 1.69171113\text{e-}01$ | $B = 9.55416151\text{e-}01$ |
| | $q = 0.003143021198831124$ | | $C = 1.98001424\text{e-}02$ |

### A.3 Demographic information of participants in the formative study

TABLE A3: Demographic information of four domain experts and 14 stakeholders participating in the formative study.

| ID | Gender | Age | Location | Market role | Participation experience Y | Co-designer | Interviwee |
|----|--------|-----|----------|-------------|----------------------------|-------------|------------|
| P1 | M | 28 | Hong Kong | Investors & Scholars/Researcher | 2 | $E_A$ | I1 |
| P2 | F | 27 | Hong Kong | Scholars/Researchers | 3 | $E_B$ | I2 |
| P3 | M | 29 | Mainland China | Investors | 4 | $E_C$ | |
| P4 | M | 25 | Hong Kong | Creator/Artists, Scholars/Researchers | 1 | $E_D$ | |
| P5 | M | 26 | Mainland China | Collectors & Brokers/Moderators | 4 | | |
| P6 | F | 28 | Mainland China | Investors | 2 | | |
| P7 | M | 30 | Australia | Creator/Artists, Collectors, Brokers/ Moderators, and Critics | 3 | | |
| P8 | F | 27 | Mainland China | Collectors & Brokers/ Moderators | 2 | | |
| P9 | M | 20 | Mainland China | Collectors, Brokers/ Moderators, Investors | 2 | | |
| P10 | M | 28 | Mainland China | Investors | 2 | | |
| P11 | F | 36 | Mainland China | Investors | 1 | | I3 |
| P12 | F | 28 | Mainland China | Collectors | 5 | | I4 |
| P13 | M | 26 | Mainland China | Collectors | 1 | | |
| P14 | M | 29 | Mainland China | Investors | 2 | | |
| P15 | F | 26 | Mainland China | Scholars/Researchers | 0.5 | | |
| P16 | F | 26 | Mainland China | Scholars/Researchers | 0.5 | | I5 |
| P17 | M | 28 | Hong Kong | Collectors | 1 | | |
| P18 | M | 26 | Mainland China | Creator/Artists, Collectors, Investors | 5 | | I6 |

### A.4 Results of the questionnaire survey in the formative study

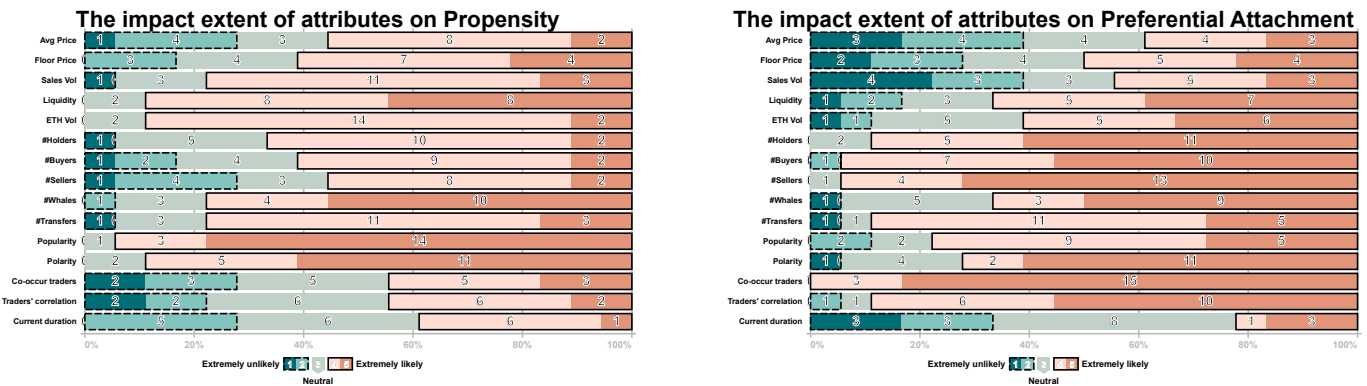

(a) Results of the survey about the impact extent of 15 potential attributes on **Propensity** mechanism.

(b) Results of the survey about the impact extent of 15 potential attributes on **Preferential Attachment** mechanism.

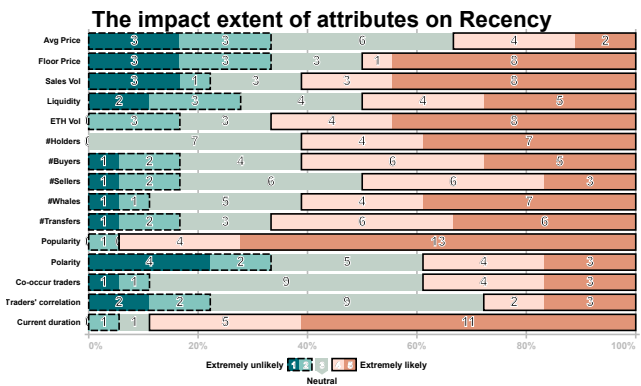

(c) Results of the survey about the impact extent of 15 potential attributes on **Recency** mechanism.

Fig. A2: Results of five-point Likert-scale questionnaire survey in the formative study. We enumerated 15 potential attributes for participants to estimate and summarized 12 fundamental impact attributes from their responses.

# APPENDIX B

TABLE B4: Detailed information of 70 sampled NFT projects, including project names, smart contracts, Twitter hashtags, and start date.

| Project Name | Smart Contract | Twitter Hashtag | Start Date |
| --- | --- | --- | --- |
| Art Gobblers | 0x60bb1e2aa1c9acafb4d34f71585d7e959f387769 | #cryptopunks | 6/1/2017 0:00 |
| - TronWars - | 0x537b2279d8f625a1b74cf3c1f0e2122fb047a6b0 | #BAYC | 4/1/2021 0:00 |
| 0N1Force | 0x3bf2922f4520a8ba0c2efc3d2a1539678dad5e9d | #MAYC | 8/29/2021 0:00 |
| 3Landers | 0xb4d06d46a8285f4ec79fd294f78a881799d8ced9 | #azuki | 1/10/2022 13:19 |
| Akutars | 0xaad35c2dadbe77f97301617d82e661776c891fa9 | #CloneX | 12/12/2021 21:50 |
| alien frens | 0x123b30e25973fecd8354dd5f41cc45a3065ef88c | #moonbirds | 4/16/2022 0:38 |
| Animetas | 0x18df6c571f6fe9283b87f910e41dc5c8b77b7da5 | #doodles | 8/21/2021 0:00 |
| Azuki | 0xed5af388653567af2f388e6224dc7c4b3241c544 | #BAKC | 6/19/2021 0:00 |
| BEANZ Official | 0x306b1ea3ecdf94ab739f1910bbda052ed4a9f949 | #coolcats | 7/1/2021 0:00 |
| Bored Ape Kennel Club | 0xba30e5f9bb24caa003e9f2f0497ad287fdf95623 | #beanz | 4/23/2022 17:13 |
| Bored Ape Yacht Club | 0xbc4ca0eda7647a8ab7c2061c2e118a18a936f13d | #wow | 7/20/2021 0:00 |
| Boss Beauties | 0xb5c747561a185a146f83cfff25bdfd2455b31ff4 | #cryptoads | 9/9/2021 0:00 |
| BoxcatPlanet Official | 0x13bd972b0bfaefc9538a43c1fda11d71c720cd47 | #pudgypenguins | 8/5/2021 0:00 |
| Call Me Cutie Pie | 0x98f832d1e7808652f380db21568dc96efc02e7da | #CyberKongz | 4/1/2021 0:00 |
| CLONE X - X TAKASHI MU-RAKAMI | 0x49c6f5d44e70224e2e23fdcdd2c053f30ada28b | #goblintown | 5/20/2022 0:16 |
| Cool Cats NFT | 0x1a92f7381b9f03921564a437210bb9396471050c | #HAPE | 1/15/2022 17:17 |
| Coolman's Universe | 0xa5c0bd78d1667c13bfb403e2a3336871396713c5 | #0n1force | 7/20/2021 0:00 |
| CrypToadz by GREMPLIN | 0x1cb1a5e65610aeff2551a50f76a87a7d3fb649c6 | #MekaVerse | 9/23/2021 0:00 |
| CryptoBatz by Ozzy Osbourne | 0xc8adfb4d437357d0a656d4e62fd9a6d22e401aa0 | #InvisibleFriends | 2/22/2022 4:48 |
| CryptoPunks | 0xb47e3cd837ddf8e4c57f05d70ab865de6e193bbb | #mfers | 11/30/2021 6:20 |
| CryptoSkulls | 0xc1caf0c19a8ac28c41fe59ba6c754e4b9bd54de9 | #karafuru | 2/4/2022 17:35 |
| CyberKongz | 0x57a204aa1042f6e66dd7730813f4024114d74f37 | #hashmask | 1/28/2021 0:00 |
| CyberKongz VX | 0x7ea3cca10668b8346aec0bf1844a49e995527c8b | #FLUFWorld | 7/22/2021 0:00 |
| DeadFellaz | 0x2acab3dea77832c09420663b0e1cb386031ba17b | #3landers | 2/18/2022 20:09 |
| DEGEN TOONZ COLLECTION | 0x19b86299c21505cdf59ce63740b240a9c822b5e4 | #Phantabear | 12/31/2021 16:11 |
| Doodles | 0x8a90cab2b38dba80c64b7734e58ee1db38b8992e | #CyberKongz | 4/1/2021 0:00 |
| Dooplicator | 0x466cfcd0525189b573e794f554b8a751279213ac | #GutterCatGang | 5/29/2021 0:00 |
| FLUF World | 0xccc441ac31f02cd96c153db6fd5fe0a2f4e6a68d | #Deadfellaz | 8/13/2021 0:00 |
| GalacticApes | 0x12d2d1bed91c24f878f37e66bd829ce7197e4d14 | #v1punks | 1/17/2022 18:15 |
| Galaxy-Eggs | 0xa08126f5e1ed91a635987071e6ff5eb2aeb67c48 | #KaijuKingz | 10/9/2021 0:00 |
| goblintown.wtf | 0xbce3781ae7ca1a5e050bd9c4c77369867ebc307e | #LazyLions | 7/31/2021 0:00 |
| Gutter Cat Gang | 0xedb61f74b0d09b2558f1eeb79b247c1f363ae452 | #veefriends | 4/15/2022 1:00 |
| HAPE Prime | 0x4db1f25d3d98600140dfc18deb7515be5bd293af | #wow | 3/26/2022 17:32 |
| Hashmasks | 0xc2c747e0f7004f9e8817db2ca4997657a7746928 | #PrimeApePlanet | 12/23/2021 4:15 |
| INNOCENTCATS NFT | 0xbabe3dec32fa870b4125b9cad1cea498dcb116c9 | #TheDogePound | 6/19/2021 0:00 |
| Invisible Friends | 0x59468516a8259058bad1ca5f8f4bff190d30e066 | #alienfrens | 12/16/2021 12:05 |
| JUNGLE FREAKS BY TROSLEY | 0x7e6bc952d4b4bd814853301bee48e99891424de0 | #psychedelic | 12/22/2021 14:15 |
| Kaiju Kingz | 0x0c2e57efddba8c768147d1fdf9176a0a6ebd5d83 | #SuperNormal | 1/25/2022 7:09 |
| Karafuru | 0xd2f668a8461d6761115daf8aeb3cdf5f40c532c6 | #VOX | 7/18/2021 0:00 |
| Lazy Lions | 0x8943c7bac1914c9a7aba750bf2b6b09fd21037e0 | #CMUnity | 12/18/2021 6:55 |
| Lives of Asuna | 0xaf615b61448691fc3e4c61ae4f015d6e77b6cca8 | #SupDucks | 6/27/2021 0:00 |
| MekaVerse | 0x9a534628b4062e123ce7ee2222ec20b86e16ca8f | #renga | 9/2/2022 7:53 |
| mfers | 0x79fcdef22feed20eddacbb2587640e45491b757f | #MurakamiFlowers | 5/1/2022 8:23 |
| Monaco Planet Yacht | 0x21bf3da0cf0f28da27169239102e26d3d46956e5 | #MoonbirdsOddities | 5/22/2022 18:34 |
| Moonbirds | 0x23581767a106ae21c074b2276d25e5c3e136a68b | #svsnft | 9/7/2021 0:00 |
| Moonbirds Oddities | 0x1792a96e5668ad7c167ab804c04c2395ce54d | #cryptoskulls | 1/12/2022 11:29 |
| Murakami.Flowers Official | 0x7d8820fa92eb1584636f4f5b8515b5476b75171a | #livesofasuna | 1/27/2022 19:24 |
| Murakami.Flowers Seed | 0x341a1c534248966c4b6afad165b98daed4b964ef | #murigang | 3/13/2022 21:51 |
| MURI by Haus | 0x4b61413d4392c806e6d0ff5ee91e6073c21d6430 | #memeland | 7/20/2022 21:19 |
| Mutant Ape Yacht Club | 0x60e4d786628fea6478f785a6d7e704777c86a7c6 | #TOONZ | 2/14/2022 10:16 |
| My Pet Hooligan | 0x09233d553058c2f42ba751c87816a8e9fae7ef10 | #junglefreaks | 9/19/2021 0:00 |
| NAH Fungible Bones | 0x0ee24c748445fb48028a74b0ccb66b46d7d3e3b33 | #GalacticApes | 3/28/2021 0:00 |

Continued on next page

**TABLE B4 – continued from previous page**

| Project Name | Smart Contract | Twitter Hashtag | Start Date |
|---|---|---|---|
| PhantaBear | 0x67d9417c9c3c250f61a83c7e8658dac487b56b09 | #tubbycats | 2/22/2022 0:45 |
| Prime Ape Planet PAP | 0x6632a9d63e142f17a668064d41a21193b49b41a0 | #quirkies | 2/9/2022 11:55 |
| Psychedelics Anonymous Genesis | 0x75e95ba5997eb235f40ecf8347cdb11f18ff640b | #mypethooligan | 12/14/2021 0:58 |
| Pudgy Penguins | 0xbd3531da5cf5857e7cfaa92426877b022e612cf8 | #bossbeauties | 8/13/2021 0:00 |
| Quirkies Originals | 0x3903d4ffaaa700b62578a66e7a67ba4cb67787f9 | #akudreams | 4/25/2022 18:45 |
| RENGA | 0x394e3d3044fc89fcdd966d3cb35ac0b32b0cda91 | #doodles | 5/13/2022 0:27 |
| Sneaky Vampire Syndicate | 0x219b8ab790decc32444a6600971c7c3718252539 | #soulda | 7/4/2022 23:59 |
| Soulda16Club | 0xe361f10965542ee57d39043c9c3972b77841f581 | #CallMeCutiePie | 6/9/2022 23:59 |
| SupDucks | 0x3fe1a4c1481c8351e91b64d5c398b159de07cbc5 | #boxcatplanet | 8/29/2022 14:22 |
| The Doge Pound | 0xf4ee95274741437636e748ddac70818b4ed7d043 | #GalaxyEggs | 8/19/2021 0:00 |
| The Potatoz | 0x39ee2c7b3cb80254225884ca001f57118c8f21b6 | #Cryptobatz | 1/16/2022 22:12 |
| tubby cats by tubby collective | 0xca7ca7bcc765f77339be2d648ba53ce9c8a262bd | #animetas | 6/22/2021 0:00 |
| V1 CryptoPunks Wrapped | 0x282bdd42f4eb70e7a9d9f40c8fea0825b7f68c5d | #monacoplanet | 11/12/2021 13:24 |
| VeeFriends Series 2 | 0x9378368ba6b85c1fba5b131b530f5f5bedf21a18 | #nahfungiblebone | 11/22/2021 23:28 |
| Vox Collectibles | 0xad9fd7cb4fc7a0fbce08d64068f60cbde22ed34c | #thetronwars | 2/3/2022 17:07 |
| World of Women | 0xe785e82358879f061bc3dcac6f0444462d4b5330 | #murakami-flowers | 3/25/2022 8:48 |
| World of Women Galaxy | 0xf61f24c2d93bf2de187546b14425bf631f28d6dc | #artgobblers | 10/31/2022 4:39 |
| Zipcy's SuperNormal | 0xd532b88607b1877fe20c181cba2550e3bbd6b31c | #innocentcats | 1/26/2022 18:15 |

TABLE B5: Introduction of data properties belonging to three sub-datasets and mapping the correspondence between datasets and mechanisms.

| Data Type | Data Properties | Mechanisms | Sources |
|---|---|---|---|
| D1: NFT Transaction Data | Smart contract address; Time stamps; From_address; To_address. | M2 | Etherscan, NFTGO |
| D2: Project Character Data | Token info, e.g., create date and Twitter_hashtags; Historical records of floor prices and sales; Number of holders; Number of unique buyers, i.e., whale accounts. | M2, M3 | Etherscan, NFTGO |
| D3: Social Media Data | Twitter content searched by Twitter_hashtags; Number of tweets searched by Twitter_hashtags; Number of retweets, replies, likes, and quotes. | M1, M3 | Twitter |

## APPENDIX C

TABLE C6: Demographic information of 13 stakeholders participating in the user study to evaluate the *informativeness*, *effectiveness*, *visual design and interaction*, and *usability* of *NFTracer*. Specifically, yr(s) stands for year(s), and min(s) stands for minute(s).

| ID | Gender | Age | Market Role | Participation Experience | Time spent on T1 | Time spent on T2 | Time spent on T3 |
|----|--------|-----|-------------|--------------------------|------------------|------------------|------------------|
| S1 | Female | 25 | Scholars/Researchers | 0.4 yr | 10 mins | 6 mins | 7 mins |
| S2 | Male | 28 | Collectors | 1 yr | 9 mins | 5 mins | 3 mins |
| S3 | Male | 27 | Scholars/Researchers; Collectors | 1 yr | 6 mins | 5 mins | 3.5 mins |
| S4 | Female | 24 | Collectors; Scholars/Researchers | 0.9 yr | 8 mins | 2 mins | 5 mins |
| S5 | Female | 27 | Collectors; Creator/Artist | 2.3 yrs | 7 mins | 3 mins | 5 mins |
| S6 | Male | 28 | Investors; Critics; Brokers | 2 yrs | 7.5 mins | 4 mins | 7 mins |
| S7 | Male | 27 | Collectors | 0.9 yr | 8 mins | 3 mins | 2 mins |
| S8 | Female | 28 | Creator/Artists | 0.7 yr | 10 mins | 4 mins | 5 mins |
| S9 | Male | 32 | Scholars/Researchers | 1 yr | 5 mins | 4 mins | 3 mins |
| S10 | Female | 28 | Scholars/Researchers | 1 yr | 7.5 mins | 7 mins | 6 mins |
| S11 | Male | 31 | Investors; Scholars/Researchers | 2 yrs | 6 mins | 5 mins | 3.5 mins |
| S12 | Female | 26 | Collectors; Scholars/Researchers | 0.6 yr | 10.5 mins | 8 mins | 9 mins |
| S13 | Female | 28 | Critics; Scholars/Researchers | 0.4 yr | 10 mins | 4 mins | 5 mins |

TABLE C7: Description of questions for seven-point Likert scale applied in the user study.

| **Informativeness** | |
|---|---|
| Q1 | The system provides significant and sufficient information for users to accomplish the three required analysis tasks. |
| Q2 | The system provides significant and sufficient information for users to freely explore and derive insights about substitutive systems accordingly. |
| Q3 | The information provided by this system is easy to access and interpret. |
| **Effectiveness** | |
| Q4 | The system is effective for users to understand, identify, and compare mechanisms governing substitutive systems in NFT marketplaces. |
| Q5 | The system is effective for users to observe the evolution of the overall substitutive systems in NFT marketplaces. |
| Q6 | The system is effective for users to identify and compare similar NFT projects with a higher probability of substituting with each other. |
| Q7 | The system is effective for users to investigate and compare the temporal impact dynamics as well as growth patterns of individual NFT projects. |
| **Visual design and interactions** | |
| Q8 | The visual design and interactions of the Propensity Analysis View are intuitive. |
| Q9 | The visual design and interactions of the Mechanisms Analysis View are intuitive. |
| Q10 | The visual design and interactions of the Substitution View are intuitive. |
| Q11 | The visual design and interactions of the Impact Dynamic View are intuitive. |
| Q12 | The multiple-view design and cross-view interactions help analyze the substitutive systems in NFT marketplaces. |
| **Usability** | |
| Q13 | It was easy to learn the system. |
| Q14 | It was easy to use the system. |
| Q15 | I would like to recommend this system to other stakeholders in NFT communities. |

TABLE C8: Examples of responses and suggestions from stakeholders who participated in the user study. Specifically, "P" represents positive sentiment, and "N" represents negative sentiment.

**Informativeness**

| ID | Quotation Examples | Sentiment |
|---|---|---|
| S1 | "I can find clear in-flow and out-flow data between NFT projects in the Substitution View, which helps a lot when comparing competitive NFT projects." | P |
| S4 | "I like the Impact Dynamic View most, specifically how it presents co-occurring stakeholders and the coefficient correlation, which gives me many details for further analysis." | P |
| S6 | "I would like more information about quantifying substitution mechanisms on the interface, for sometimes I would forget their definitions." | N |

**Effectiveness**

| ID | Quotation Examples | Sentiment |
|---|---|---|
| S3 | "The design of the Substitution View is novel, and I also find the interactions interesting and diverse." | P |
| S12 | "I think the Mechanisms Analysis View is the most effective because it helps me characterize and compare different groups quickly." | P |
| S4 | "The set matrix in the Preferential Attachment View is impressive [...] The glyph design is harmonious and effective, which increases my interest in exploration." | P |

**Visual design and interaction**

| ID | Quotation Examples | Sentiment |
|---|---|---|
| S2 | "I think it would be better to expand the node-link graph in the Substitution View a little bit more, and thus, I can compare the amount of migrated stakeholders more quickly." | N |
| S12 | "I enjoy the color palette of the interface. It illustrated a great sense of aesthetics." | P |
| S10 | "I really appreciate the design of the Mechanisms Analysis View and the upper part of the Impact Dynamic View, which is straightforward and intuitive." | P |

**Usability**

| ID | Quotation Examples | Sentiment |
|---|---|---|
| S2 | "I can establish an analysis framework through the three mechanisms [...] Actually, I didn't realize that recency is so influential when analyzing the impact dynamics of NFT projects." | P |
| S10 | "I appreciate the straightforward tabular data in the Propensity Analysis View, which reduces my learning time and also helps me compare different NFT projects." | P |
| S11 | "This system is particularly beneficial for brokers and moderators in NFT communities, as they can use it to track stakeholders' behaviors." | P |